\documentclass[aip,jrse,showpacs,onecolumn]{revtex4}
\usepackage{graphicx}
\usepackage{dcolumn}
\usepackage{amssymb,amsmath}
\usepackage{natbib}
\usepackage{subfigure}
\usepackage{epstopdf}
\usepackage{placeins}
\usepackage{color}
\usepackage{epstopdf}
\usepackage{chngpage}

\begin{document}

\noindent Copyright 2014 American Institute of Physics. This article may be downloaded for personal use only. Any other use requires prior permission of the author and the American Institute of Physics. The following article appeared in J. of Renewable and Sustainable Energy 6, 043102 (2014), and may be found at http://dx.doi.org/10.1063/1.4885114.

\title{Temporal structure of aggregate power fluctuations in large-eddy simulations of extended wind-farms}
\author{Richard J. A. M. Stevens$^{1,2}$ and Charles Meneveau$^{1}$}
\affiliation{
$^1$Department of Mechanical Engineering \& Center for Environmental and Applied Fluid Mechanics, Johns Hopkins University, Baltimore, Maryland 21218, USA\\
$^2$Department of Physics, Mesa+ Institute, and J.\ M.\ Burgers Centre for Fluid Dynamics, University of Twente, 7500 AE Enschede, The Netherlands}

\date{\today}

\begin{abstract}
Fluctuations represent a major challenge for the incorporation of electric power from large wind-farms into power grids. 
Wind farm power output fluctuates strongly in time, over various time scales. Understanding these fluctuations, especially their spatio-temporal characteristics, is 
particularly important for the design of backup power systems that must be readily available in conjunction with wind-farms. 
In this work we analyze the power fluctuations associated with the wind-input variability at scales between minutes to several hours, using large eddy simulations (LES) of extended wind-parks, interacting with the atmospheric boundary layer. LES studies enable careful control of parameters and availability of wind-velocities simultaneously across the entire wind-farm. The present study focuses on neutral atmospheric conditions and flat terrain, using actuator-disk representations of the individual wind-turbines. We consider power from various aggregates of wind-turbines such as the total average power signal, or signals from sub-averages within the wind-farm. Non-trivial correlations are observed due to the complex interactions between turbines placed downstream of each other, and they lead to noticeable spectral peaks at frequencies associated with the inter-turbine spacings when the wind-direction is completely fixed. In that case we observe that the frequency spectra of the total wind-farm output show a decay that follows approximately a $-5/3$ power-law scaling regime, qualitatively consistent with some observations made in field-scale operational wind-parks (Apt, 2007). We find that these features are still observed when the wind-speed varies in magnitude. However, significant changes in the wind-direction over time tend to smooth out the observed spectral peak and reduce the extent of the observed $-5/3$ power-law.
\end{abstract}

\pacs{47.85.Np,47.11.-j,47.27.ep,47.27.nb,47.27.te}

\maketitle

\section{Introduction} \label{section1}
In 2013 almost $3\%$ of the global electricity demand came from wind-turbines \cite{WIND2013} and various scenarios \cite{eu2007,us2008} aim for a contribution up to $20 \%$ by $2030$. Several countries have already achieved a relatively high usage of wind-power in 2013, such as $30\%$ in Denmark, and $20\%$ in Portugal and $18\%$ Spain \cite{WIND2013}. However, to realize the targets worldwide, larger onshore and offshore wind-farms covering increasingly larger surface areas will be required.

Much work has already been done on the optimization of single wind-turbines \cite{sne98,bur01,her07} and a significant number of studies have focused on the structure of wakes from individual wind-turbines \cite{cre96,wha00,iva98,ebe99,mag99,ver03,sor02b,med06,cha09,cha10}. Also, models that deal with superposition of wakes stemming from a finite number of wind-turbines (see e.g.\ Ref. \cite{bar07} for an overview of different models) have been developed and some even consider the limit of infinitely many wind-turbines, as done in pioneering works by Lissaman \cite{lis79} and Frandsen \cite{fra92}. As the characteristic height of the atmospheric boundary layer (ABL) is about 1 km, large wind-farms can be considered to approach the limit of an infinite wind-farm when their dimensions exceed 10-20 km \cite{cal10}.

When the problem of large wind parks is approached from the side of large atmospheric scales, wind-turbine arrays are often modeled as surface roughness elements or via net drag coefficients, associated to an increased roughness length that needs to be parameterized. This approach is necessary for simulations in which the effect of large wind-farms at regional and global scales is considered. Examples are studies that aim at predicting the effect of large wind-farms on the global climate \cite{kei04,wan10}, regional meteorology \cite{bai04}, or short time weather patterns \cite{bar10,zho12}. In such simulations, the horizontal computational resolution near the ground is often significantly coarser than the height of the ABL and therefore insufficient to study the physical mechanisms that are important in wind-farms. Due to the large separation of scales when modeling a large wind-farm, many studies \cite{san11,ver03} have used RANS (Reynolds-averaged Navier-Stokes) to model large wind parks. However, with the aim of capturing temporal variability and fluctuations, a more detailed --- unsteady ---  simulation framework is required. 

Recently, large eddy simulations (LES) have been applied to study the interaction between wind-turbines and the turbulent ABL \cite{jim07,cal10,mey10,iva10,san11,chu12,wu13}. Only a limited number of LES simulations focused on large wind-turbine parks. Ivanell \cite{iva10} performed LES of the Horns Rev wind-plant and assumed periodic conditions on the side boundaries to approximate the full plant aerodynamics, while Churchfield et al.\ \cite{chu12} used LES to model the Lillgrund wind-plant. Meyers and Meneveau \cite{mey10} and Calaf et al.\ \cite{cal10,cal11} performed LES in a horizontally periodic domain to study the effect of wind-turbines spacing on the total power output of an infinitely large wind-farm and the associated momentum and kinetic energy transport. They showed that in large wind-farms the total power output is mainly determined by the vertical fluxes of kinetic energy in the wind-farm, which was also observed in the wind-tunnel experiments of Cal et al.\ \cite{cal10b}. An analysis of the scalar transport \cite{cal11} showed that the presence of wind-farms increases the scalar transport from the bottom surface by only small amounts. Wu and Port\'e-Agel \cite{wu13} showed that inside staggered farms, the relatively longer separation between consecutive downwind-turbines allows the wakes to recover more compared to the aligned configuration although it has been shown by Stevens et al.\ \cite{ste14b} that the staggered arrangement is not necessarily optimal. These simulations are in agreement with experiments from Chamorro et al.\ \cite{cha11b} which show a higher vertical momentum transfer into the hub-height plane for a staggered wind-farm than for an aligned wind-farm. Yang et al.\ \cite{yan12} showed with simulations that in infinitely large aligned wind-farms the stream-wise spacing is more important for the average power production than the span-wise spacing. Most of the work described above focuses on the mean properties of power produced by large wind-farms. However, there are huge temporal fluctuations in the flow speeds, which lead to strong variations of the power output over time for individual wind-turbines and for the entire farm (see e.g. Fig. 6 from Ref. \cite{cal10}).

Detailed knowledge about spectral characteristics  of power output variations will allow a more efficient match between the natural variations of the wind-park output power and the fill-in power that has to be provided \cite{rug12,gay12}.  The power fluctuations of the turbines depend on turbine regulation and electro-mechanical control systems, as well as on the wind-input to individual turbines in the array. The latter includes wind-velocity fluctuations from the effect of the wakes created by upstream turbines, on the operation of downstream turbines, as well as on inter-turbine correlations. In addition, the inherent unsteadiness of the atmospheric boundary layer (ABL) and its interactions with the turbines are important. Based on earlier field measurements \cite{thi96,thi01,thi06,ama07} and wind-tunnel experiments by Chamorro and Port\'e-Agel \cite{cha09,cha10} it is known that the spectrum reveals a signature of the helicoidal tip vortices shed by the turbine blades, i.e.\ a local peak in the spectrum at a frequency coincident with three times the frequency of the turbine. This effect has been modeled by S$\o$rensen et al.\ \cite{sor02} and has also been observed in LES \cite{lee12,chu12b,sim12}. These studies mainly focus on events from seconds to minutes, while less studies focus on spectra obtained over longer timescales. Wind-tunnel experiments \cite{cha10,zha13,zha12b} and LES \cite{chu12b} have shown that the velocity spectra in the wakes of turbines have a $-5/3$ Kolmogorov frequency spectrum that transitions towards the $-1$ range for lower frequencies. Data from field measurements for this low frequency range are limited and do not give information on the wind-park design. An additional difficulty is that measurement of low frequencies in field experiments are always influenced by changes in the atmospheric conditions, which makes direct comparison among data difficult \cite{apt07,ama07,sor07,kat10,tar11}. We note that there are also some studies that consider the relation in power output of geographically separated wind-parks \cite{apt07,kat10,tar11} and efforts are undertaken to model and quantify these phenomena \cite{bou12,ros12}.

In this study we use LES to characterize the frequency spectra of the power output of extended wind-farms for temporal scales ranging from about 1 minute to several hours. A number of simplifying assumptions are used. First, we focus on a relatively simple regime of neutral atmospheric stability. Second, we include only temporal variability arising from the micro-scale ABL dynamics and do not model long term weather phenomena. Finally, we represent the turbines using a simple averaged actuator disk description, without modeling the structural response, electromechanical control systems, coupling to electric grid, etc. These simplifications enable us to simulate extended wind-farms with many turbines and thus allow us to study the interaction between the turbines on extended spatial scales, and perform very long simulations that cover many hours in duration. In order to characterize fluctuations across the spatial extent of the wind farm, we consider various wind-turbine aggregates, such as the average power signal from the entire domain, or specific sub-averages in various directions within the wind-farm. The fluctuations of these sub-averages are expected to depend critically upon wind turbine interactions and wakes. 
We start with an introduction to the computational methods that are applied, before we discuss the results of the simulations. 

\section{LES wind farm model} \label{section2}
In this work we consider flow that is not thermally stratified. Therefore, the LES is based on the filtered incompressible Navier-Stokes equations for neutral flows and the continuity equation, i.e.\,
\begin{eqnarray}
\partial_{t} \tilde{u}_i + \partial_{j} (\tilde{u}_{i} \tilde{u}_{j}) &=& - \partial_{i} \tilde{p}^{*} - \partial_{j} \tau_{ij} -  \partial_{i} p_{\infty}/ \rho + f_{i} \\
\partial_{i} \tilde{u}_{i} &=& 0
\end{eqnarray}
where $\tilde{u}$ is the filtered velocity field and $\tilde{p}^*$ is the filtered modified pressure equal to $ \tilde{p}/\rho + \tau_{kk}/3-p_\infty/\rho$. Further, $\tau_{ij}$ is the subgrid-scale stress term. Its deviatoric part ($\tau_{ij} -\delta_{ij} \tau_{kk}/3$) is modeled using an eddy viscosity subgrid-scale model, as discussed further below; the trace of this term ($\tau_{kk} / 3$) is combined into the modified pressure, as is common practice in LES of incompressible flow. The force $f_{i}$ is added for modeling the effects of the wind-turbines in the momentum equation.
Since simulations are done at very large Reynolds numbers and the bottom surface as well as the wind-turbine effects are parametrized, viscous stresses are neglected. We use an imposed pressure gradient
\begin{equation}
\label{eq_forcing1}
-  {\rho}^{-1} ~\partial_1 p_{\infty} ~=  G ~\left(1+A \sin \left[2 \pi  {t}/{t_{p1}}\right]\right) \cos \left(\frac{\pi}{180} \left(\phi+ \mu \sin \left[2 \pi  {t}/{t_{p2}}\right]\right)\right) 
\end{equation}
in the $x_1$ direction and 
\begin{equation}
\label{eq_forcing2}
-  {\rho}^{-1} ~\partial_2 p_{\infty} ~= G ~  \left(1+A \sin \left[2 \pi  {t}/{t_{p1}}\right]\right) \sin \left(\frac{\pi}{180} \left(\phi + \mu \sin \left[2 \pi  {t}/{t_{p2}}\right]\right)\right) 
\end{equation}
in the $x_2$ direction. $A$ indicates the variation of the amplitude of the pressure forcing, which leads to a variation in the wind-speed (see section \ref{section4a}), $\phi$ indicates the average wind-direction (see section \ref{section_new2}), and $\mu$ indicates the amplitude of the directional change in the pressure forcing, which leads to a dynamically changing wind-direction (see section \ref{section_new3}).   As default values (and unless specified otherwise) we use $A=0$, $\phi=0$ and $\mu=0$, which means that the pressure forcing simplifies to $\partial_{1}p_\infty$ in the $x_1$ direction and no pressure forcing in the $x_2$ direction.  The imposed average pressure gradient magnitude $G$ selects a velocity scale given by $u_*=\sqrt{H G}$, where $H=1000$ m is the domain height used in the LES. Velocities in the simulation are dimensionless, in units of $u_*$, and length-scales are in units of $H$. Later on, when studying  temporal changes, we use $t_{p1}=15$ and $t_{p1}=20$ (see Eqs. \ref{eq_forcing1} and \ref{eq_forcing2}) in non-dimensional time-units ($H/u_*$), which corresponds to forcing periods of roughly $4.5$ and $6$ hours, respectively. The subgrid scales are modeled with the dynamic Lagrangian scale-dependent Smagorinsky model \cite{bou05}. The top boundary uses a zero vertical velocity and a zero shear stress boundary condition. At the bottom surface a classic imposed wall stress boundary condition relates the wall stress to the velocity at the first gridpoint using the standard log (Monin-Obukhov) similarity law \cite{moe84}, see also Bou-Zeid et al.\ \cite{bou05} for more details. The surface roughness ($z_{0,lo}=10^{-4}H$) is kept constant, and $\kappa=0.4$ is the Von K\'arm\'an constant. Additional details about the code are provided in Refs. \cite{por00,bou05}. 

The wind-turbines are modeled through an actuator disk approach \cite{jim07,cal10,cal11,ste13,ste14}. This approach has already been used in past studies and its advantages and drawbacks have been documented in the detailed comparisons with wind-tunnel data presented in Wu and Port\'e-Agel \cite{wu11}. They show that except for the near-wake region, the drag disk approach yields a good degree of accuracy. In particular, the low-frequency characteristics of wakes impinging upon downstream turbines are
expected to be represented accurately with this approach since downstream turbines will not be placed in the near-wake region. In the present study, the actuator disk approach is 
preferable over more detailed approaches (such as actuator line or fully resolving wind turbine blades) since it enables simulation of very large wind farms for extended periods of time. The method is based on a drag force ($F_t$) acting in the stream-wise direction according to
\begin{equation}
F_t = - \frac{1}{2} \rho C_T U^2_\infty A,
\end{equation}
where $C_T$ is the thrust coefficient, $D$ is the rotor diameter, $A=\pi D^2/4$ is the disk area of the turbine, and $U_\infty$ is an upstream (unperturbed) velocity. This is a good approach when one is modeling a single wind-turbine and there are no other interacting bodies in the numerical domain that can make specification of $U_\infty$ ambiguous. When modeling wind-farms, it is impossible to define an unperturbed upstream mean velocity since the upstream values are always affected by other upstream wind-turbines. It is thus more convenient to use the local velocity at the rotor disk $U_D$ \cite{mey10,cal10}. It can be related to an equivalent upstream unperturbed velocity through the actuator-disk theory
\begin{equation}
U_\infty = \frac{U_D}{1-a}
\end{equation}
where $a$ is the induction factor. Moreover, modeling the thrust forces acting on the fluid due to its interaction with the rotating blades requires the use of an average disk velocity. This approach assumes that the turbine is rotating at an angular velocity which is optimally adapted to the time-averaged incoming velocity, a constant tip speed-ratio, and that the ratio of lift and drag coefficient of the blades remains constant over the considered time interval, which assumes an active pitch control of the turbine that adapts the angle of the blades to changes in incoming velocity \cite{mey10}. The disk averaged velocity $U_D^T$ is evaluated from LES by averaging over the disk region, and using a first order relaxation method with a typical time of $\approx4.8$ seconds to average over time, yielding a disk averaged velocity 
\begin{equation}
U_D^T=\langle \overline{u}^{T} \rangle_D,
\end{equation}
where the subscript $D$ denotes averaging over the turbine disk region and the superscript $T$ denotes time filtering.
Then, the total thrust force can be written as
\begin{equation}
F_t = - \frac{1}{2} \rho C_T^{\prime}{\left(U_D^T\right)}^2 A,
\end{equation}
with
\begin{equation}
C_T^{\prime} =\frac{C_T}{(1-a)^2}.
\end{equation}
For the Betz limit (i.e.\, $C_T=8/9$, and $a=1/3$), we obtain $C_T^{\prime}=2$. In this study we use values which may be found in existing wind-turbines \cite{bur01} and prior LES studies \cite{jim07}, i.e.\ we use $C_T=3/4$ and $a=1/4$, which leads to $C_T^{\prime}=4/3$. Note that the value of $C_T^{\prime}$ is related to the lift and drag coefficient of the turbine blades \cite{mey10,mey12} and is less sensitive to wind-farm parameters such as the average turbine spacing. In order to obtain the applied force $f_{i}$ for LES (in which we apply the force in the mean flow direction)
the total thrust force $F_t$ for the particular wind turbine in question is divided by the  grid volume and weighted by the fraction of a grid volume intersecting the wind turbine disk (for more details, see Ref. \cite{cal10}). 

We note in passing that for wind  farm analysis it is often useful to express the thrust in relation to the average land surface area $S$ per turbine ($S=l_xl_y$, where $l_x$ and $l_y$ are the average turbine spacing in the stream-wise and span-wise direction, respectively) leading to $F_t = \frac{1}{2} \rho c_{ft}^{\prime} {U_D^T}^2 S$, with a friction coefficient $c_{ft}^{\prime}$ defined as $ c_{ft}^{\prime}={\pi C_T^{\prime}}/({4 s_x s_y})$, where $s_x=l_x/D$, and $s_y=l_y/D$. 

In this work, we will be particularly interested in the  power extracted by the turbines from the resolved flow. Based on actuator disc theory, the power  is given by \cite{mey10,mey12}
\begin{equation}
P = - F_t {U_D^T} = \frac{1}{2} \rho C_T^{\prime} {\left(U_D^T\right)}^3 A .
\end{equation}
Note that this power is not equivalent to the power on the turbine axis, which relates the torque and the rotational velocity of the turbine. The drag forces on the turbine blades increase the trust but reduce the torque, so from an energetic point of view the drag forces lead to losses, i.e.\ the mean flow energy of the ABL is converted into turbulent motion and heat. By using the power coefficients $C_P$ and $C_P^{\prime}$ (respectively with respect to $U_\infty$ and ${U_D^T}$, i.e. $C_P=C_P^{\prime}(1-a)^3$) the power on the turbine axis corresponds to
\begin{equation}
P_{ax} = P ~\frac{C_P^{\prime}}{C_T^{\prime}} .
\end{equation}
For the present simulation (mimicking operations of typical wind-turbines) we use $a\approx 1/4$. Furthermore, let us assume a power coefficient of $C_P\approx 0.34$, 
which means that $C_P^{\prime}=0.34 (1-0.25)^{-3} \approx 0.8$. Together with $C_T^\prime =4/3$ this gives $P_{ax} \approx 0.6 P$, where $P$ is given by Eq. \ref{eq_power}. Taking the disk-averaged velocity from LES at any given turbine, ${U_D^T}(t)$ as function of time, we therefore evaluate the axial power signal according to 
\begin{equation}
\label{eq_power}
P_{ax}(t) = 0.6  ~\frac{1}{2} \rho ~C_{T}^{\prime}~{\left(U_D^T\right)}^3(t)~ \frac{\pi}{4} D^2.
\end{equation}
In reality, the ratio between $C_P^{\prime}$ and $C_T^{\prime}$ depends on the turbine working regime. In operating regime II the ratio is close to optimal and in this regime the power output is not restricted, and wind-turbines work close to their aerodynamical optimal operating conditions. In this paper we assume that the turbines are always operating in this regime. Regime I, in which the aerodynamic forces cannot overcome the turbineÕs internal friction losses, and regime III, in which the turbine power is controlled at a constant level, independent of the wind-speed by stalling the turbine blades or feathering the turbine, are not considered in this paper \cite{mey10,mey12}.

In order to present the results in physical units we need to specify the velocity scale $u_*$ in m/s. This velocity scale is determined by assuming that the mean velocity at $1$km height is 12.5 m/s. That is to say, we assume a common imposed wind velocity sufficiently far above the 
wind farm.  For the periodic cases, which are performed in domains of $1$km height, $u_*$ is set by measuring the dimensionless mean velocity $[\langle \overline{u} \rangle /u_*]$ from the simulation at the top of the domain, and setting 
$u_* = 12.5~ [ \langle \overline{u} \rangle /u_*]^{-1}$ m/s. The finite size wind-farm simulations are described in more detail in
section \ref{section4b}). Briefly, they include a separate inflow domain simulation and are performed in a $2$ km high domain. For these simulations, the numerical value of $u_*$  is set again by assuming that there is a mean velocity of 12.5 m/s at a height of 1000m  
in the incoming flow that is unperturbed by the wind farm. The standard log-law $\langle \overline{u}\rangle = u_*/\kappa ~ \ln(z/z_0)$ is well approximated at that height and thus we set $u_* = 12.5 \kappa / \ln(10^4)$ m/s, since we have specified $z_0=0.1$m \
at the bottom surface (we use $\kappa = 0.4$).  The $u_*$ values determined in this way are given in tables \ref{table1}, \ref{table4}, and table \ref{table3}. For the periodic cases $u_*=0.93\pm0.05$ m/s and for the finite size windfarm $u_*\approx0.54$ m/s. The power is obtained from equation \ref{eq_power}, using $\rho=1.23$ kg m$^{-3}$ (air density at $15$ degree Celsius), $C_{T}^{\prime} = 4/3$, and $D=100$m.

\section{Baseline results: infinite wind farm with constant wind} \label{section4}

In this section we  discuss the results from the infinite wind-farm simulations. We start with a discussion on the spectra for single turbines before we discuss the collective behavior. In subsequent sections, we will discuss the effect of changing wind-speed and wind-direction before we will compare these results with the behavior found in LES of finite size wind-farms.
 
To model infinitely large wind-farms we use periodic boundary conditions in the stream-wise and span-wise direction, where we vary the resolution (given by the parameters $L_x$, $L_y$, $H$ and $N_x$, $N_y$, $N_z$) and the number of wind-turbines $N_t$. The distances among the wind-turbines ($s_x D$ and $s_yD$), where $D$ is the diameter of the turbine blades (in our case $100$ meters), and the relative sparseness of their distribution $S/A=4s_xs_y/ \pi$ is held constant. For these values we took the reference case considered by Calaf et al.\ \cite{cal10}, namely $s_x / s_y=1.5$ and $4s_xs_y / \pi= 52.36$.

\begingroup
\squeezetable
\begin{table}[h]
\caption{Summary of the simulations of the infinite wind-farms performed on different grids. The columns from left to right indicate the name of the case that is considered, the stream-wise ($L_x$) and span-wise ($L_y$) size of the domain, the resolution in stream-wise ($N_x$), span-wise ($N_y$), and vertical ($N_z$) direction, the number of wind-turbines in stream-wise and span-wise direction and whether the turbines are placed in an aligned or staggered arrangement. The next five columns give the time averaged turbine power output $\langle P \rangle$ and the standard deviation of the power output of a single wind-turbine $s_T$, all turbines $s_A$, a row of turbines $s_R$, and a column of turbines $s_C$. The last column gives the $u_*$ value for the different cases.}
\label{table1}
\begin{center}
 \begin{tabular}{|c|c|c|c|c|c|c|c|c|c|c|c|}
 \hline
 Case	& $L_x (km)$	& $L_y (km)$	&	$N_x	 \times N_y 	\times N_z$		&	$N_R \times N_C$	&	Positioning 	& $\langle P\rangle  [MW]$	& $s_T [MW]$ 	& $s_A [MW]$	& $s_R [MW]$	& $s_C [MW]$	& $u_{*} [m/s]$ \\ \hline
 A1		& $\pi$		& $\pi$		&	$64		\times 64		\times 64$		&	$ 4 \times 6$ 		&	aligned		& 1.374		& 0.584		& 0.116		& 0.211		& 0.392		& 0.949 \\ \hline		
 C1		& $2\pi$		& $\pi$		&	$128		\times 64		\times 64$		&	$ 8 \times 6$		&	aligned		& 1.365		& 0.576		& 0.099		& 0.218		& 0.287		& 0.944 \\ \hline		
 E1		& $4\pi$		& $\pi$		&	$256		\times 64		\times 64$		&	$ 16 \times 6$		&	aligned		& 1.291		& 0.546		& 0.051		& 0.207		& 0.184		& 0.927 \\ \hline		
 H1		& $8\pi$		& $\pi$		&	$512		\times 64		\times 64$		&	$ 32 \times 6$		&	aligned		& 1.323		& 0.564		& 0.044		& 0.209		& 0.144		& 0.935 \\ \hline		
 J1		& $16\pi$		& $\pi$		&	$1024	\times 64		\times 64$		&	$ 64 \times 6$		&	aligned		& 1.308		& 0.554		& 0.030		& 0.211		& 0.096		& 0.931 \\ \hline		
 A2		& $\pi$		& $\pi$		&	$128		\times 128		\times 128$	&	$ 4 \times 6$		&	aligned		& 1.252		& 0.550		& 0.117		& 0.206		& 0.368		& 0.940 \\ \hline		
 C2		& $2\pi$		& $\pi$		&	$256		\times 128		\times 128$	&	$ 8 \times 6$		&	aligned		& 1.150		& 0.499		& 0.087		& 0.190		& 0.252		& 0.913 \\ \hline		
 E2		& $4\pi$		& $\pi$		&	$512		\times 128		\times 128$	&	$ 16 \times 6$		&	aligned		& 1.110		& 0.473		& 0.054		& 0.185		& 0.163 		& 0.903 \\ \hline		
 A3		& $\pi$		& $\pi$		&	$256		\times 256		\times 256$	&	$ 4 \times 6$		&	aligned		& 1.150		& 0.504		& 0.131		& 0.200		& 0.347		& 0.918 \\ \hline		
 B1		& $\pi$		& $\pi$		&	$64		\times 64		\times 64$		&	$ 4 \times 6$ 		&	staggered		& 1.539		& 0.662		& 0.118		& 0.212		& 0.358		& 0.965 \\ \hline		 
 D1		& $2\pi$		& $\pi$		&	$128		\times 64		\times 64$		&	$ 8 \times 6$		&	staggered		& 1.478		& 0.635		& 0.104		& 0.216		& 0.235 		& 0.947 \\ \hline		
 F1		& $4\pi$		& $\pi$		&	$256		\times 64		\times 64$		&	$ 16 \times 6$		&	staggered		& 1.451		& 0.620		& 0.063		& 0.211		& 0.160		& 0.943 \\ \hline		
 B2		& $\pi$		& $\pi$		&	$128		\times 128		\times 128$	&	$ 4 \times 6$		&	staggered		& 1.332		& 0.563		& 0.108		& 0.191		& 0.295		& 0.942\\ \hline		
 D2		& $2\pi$		& $\pi$		&	$256		\times 12		\times 128$	&	$ 8 \times 6$		&	staggered		& 1.140		& 0.484		& 0.079		& 0.170		& 0.178		& 0.895 \\ \hline		
 F2		& $4\pi$		& $\pi$		&	$512		\times 128		\times 128$	&	$ 16 \times 6$		&	staggered		& 1.185		& 0.492		& 0.052		& 0.174		& 0.129		& 0.907 \\ \hline		
 \end{tabular}
\end{center}
\end{table}
\endgroup											

\begin{figure*}
\subfigure{\includegraphics[width=0.80\textwidth]{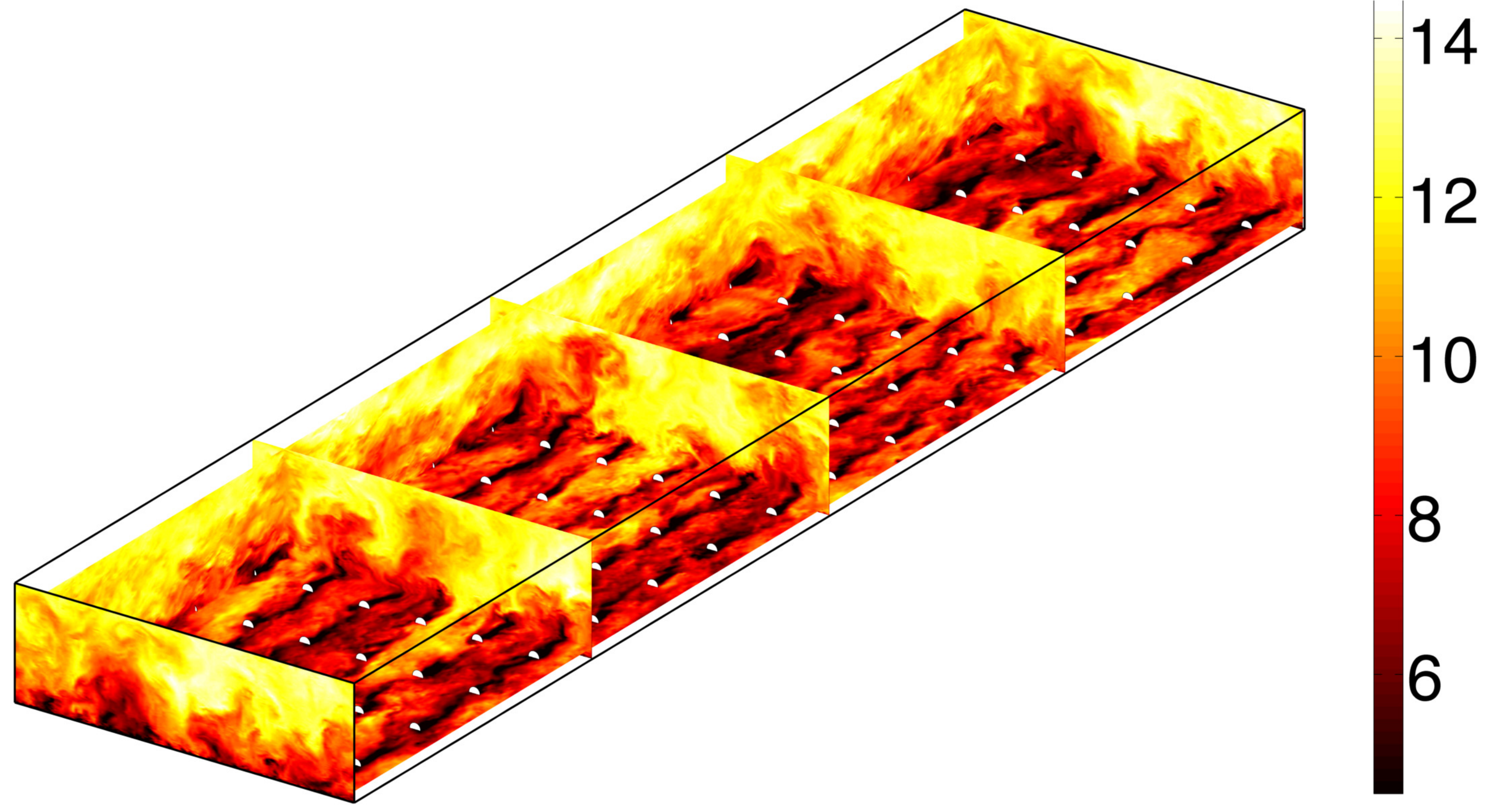}}
 \caption{Instantaneous stream-wise velocity contours (in m/s) on various planes from LES (case E2 with $16\times 6 = 96$ wind-turbines) in a three-dimensional view. The upper half of the wind-turbines is visible as white disks.}
\label{figure1}
\end{figure*}

The details for the simulations are summarized in table \ref{table1}. The simulations on the coarsest grids are indicated by $*1$ in the table. These simulations are run for about $90$ non-dimensional time units (equivalent to about $27$ hours) and the first $18$ dimensionless time units (approximately $5.4$ hours) are discarded in order to make sure that the simulation has reached its statistical stationary state. The simulations on the medium mesh, indicated by $*2$, are started from an interpolated flow solution obtained from the $*1$ simulations and have been run for similar time periods. The very long simulation times we use are necessary to calculate the spectra over a sufficiently long frequency range. We have also performed a simulation on a fine, i.e.\ $*3$ mesh for one of the cases in order to check the influence of grid resolution. We find that even the results obtained in coarsest resolution simulations capture the lower frequencies we are interested in well although there are some differences in the high frequency range. 

\begin{figure}
\subfigure{\includegraphics[width=0.47\textwidth]{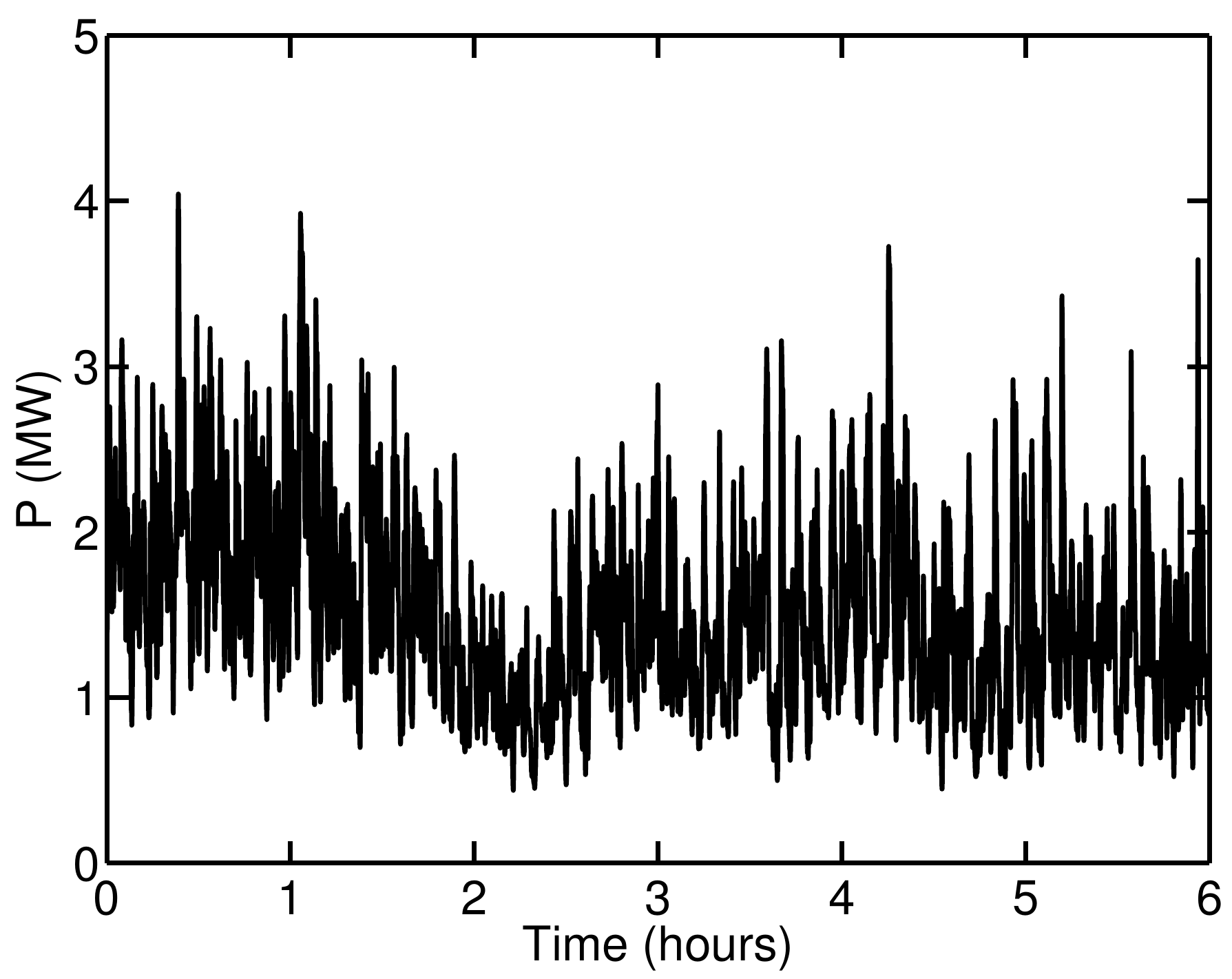}}
\caption{Power output of a turbine over a $6$ hour period in simulation A2.}
\label{figure2}
\end{figure}

\begin{figure*}
\subfigure[]{\includegraphics[width=0.47\textwidth]{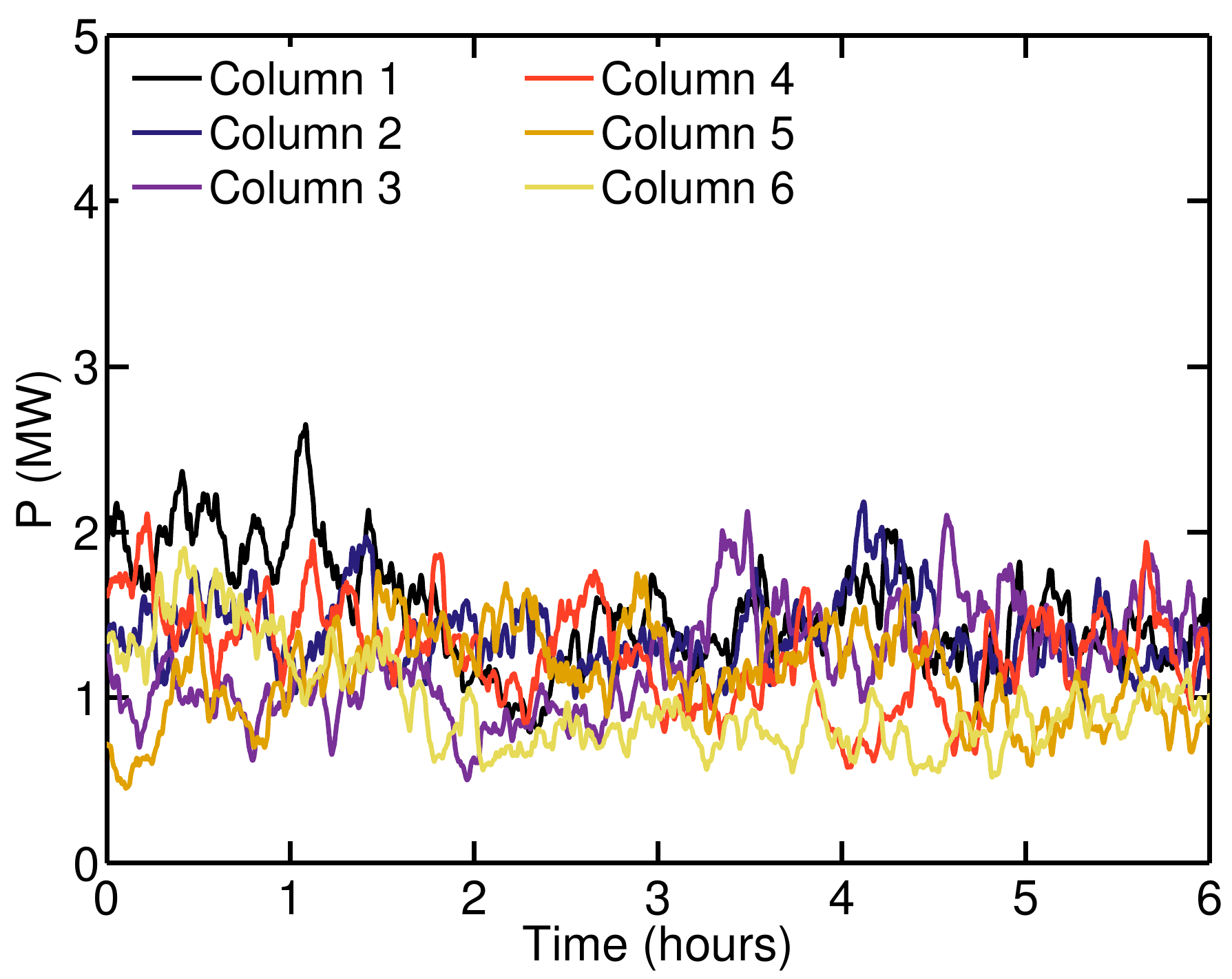}}
\subfigure[]{\includegraphics[width=0.47\textwidth]{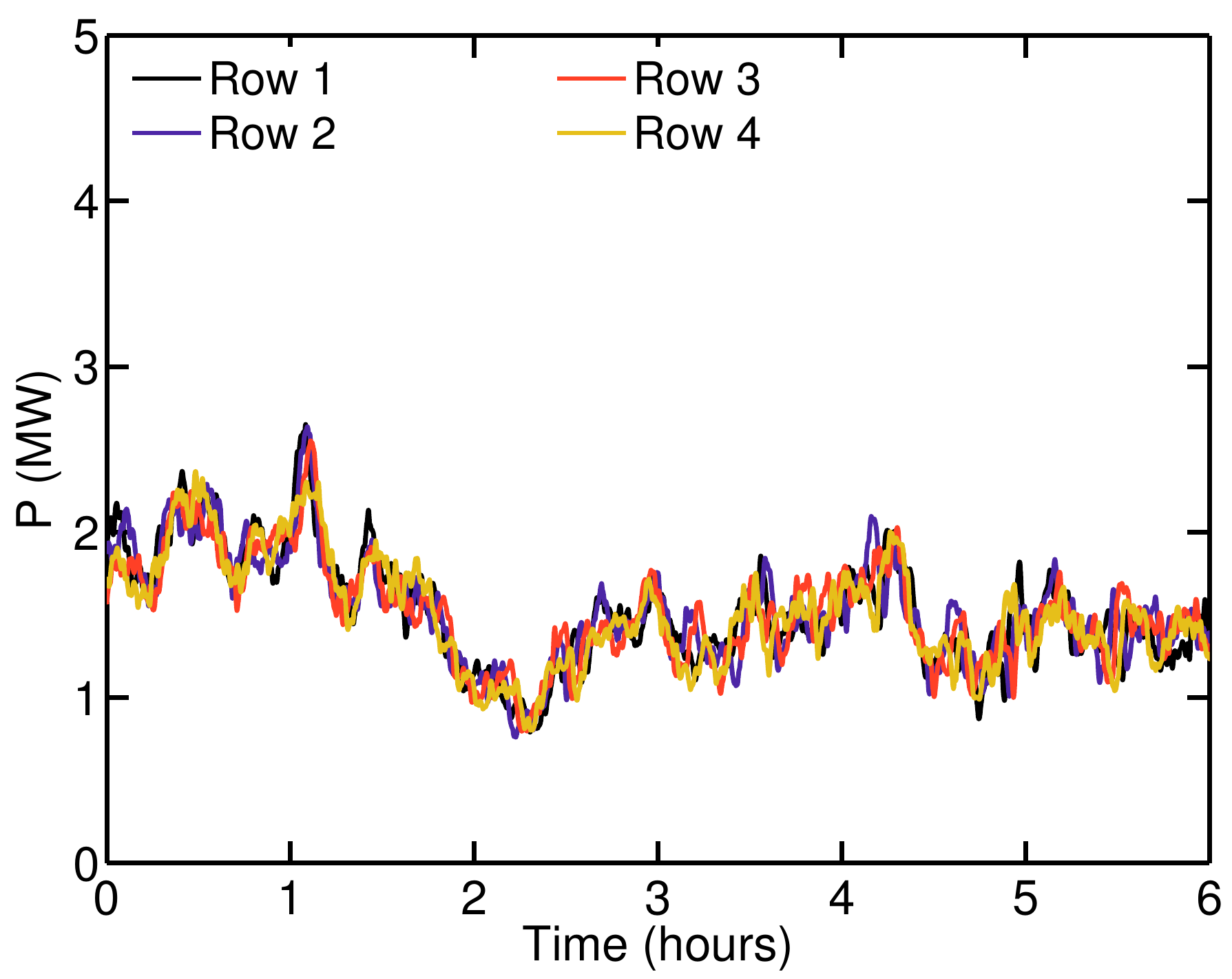}}
\subfigure[]{\includegraphics[width=0.47\textwidth]{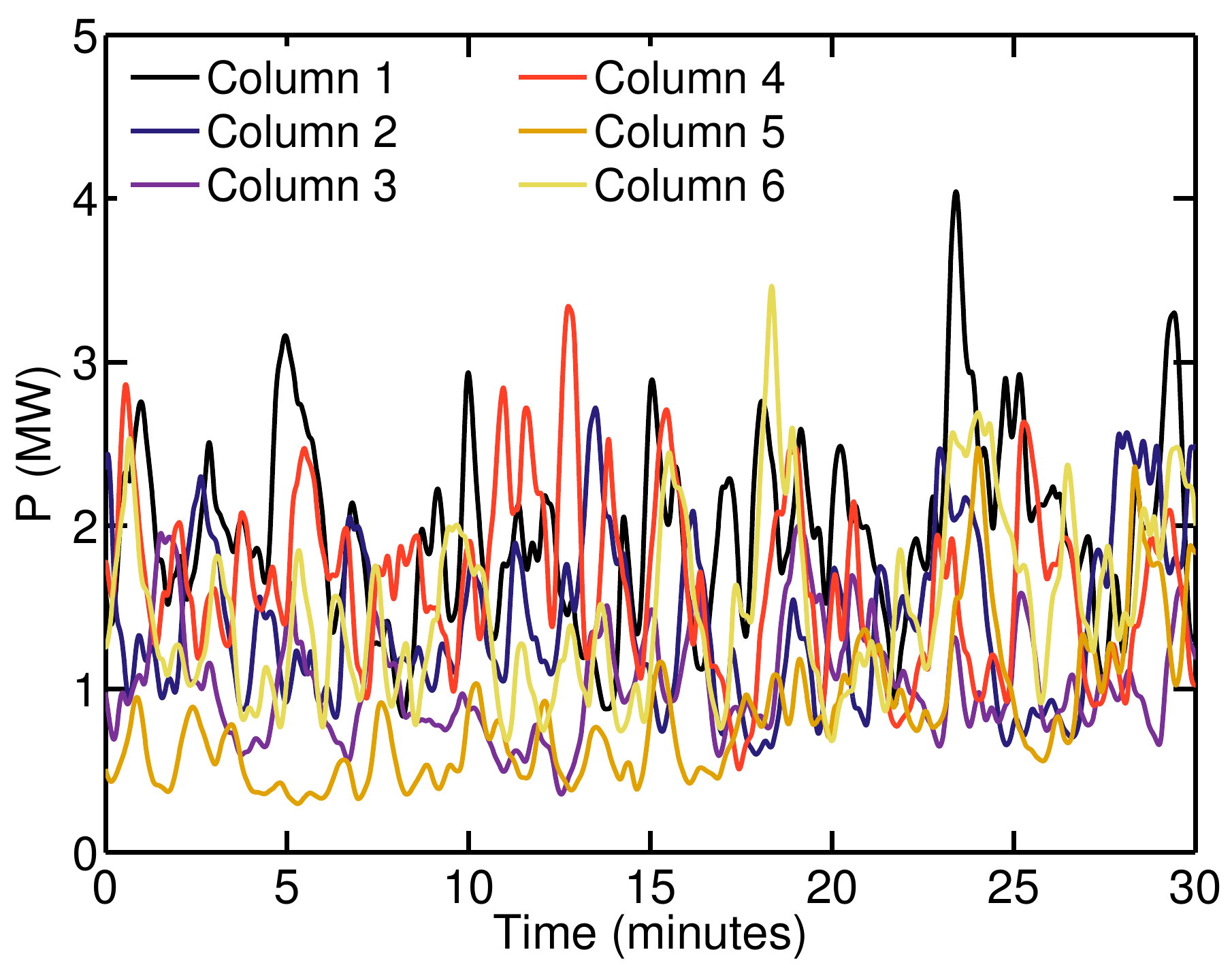}}
\subfigure[]{\includegraphics[width=0.47\textwidth]{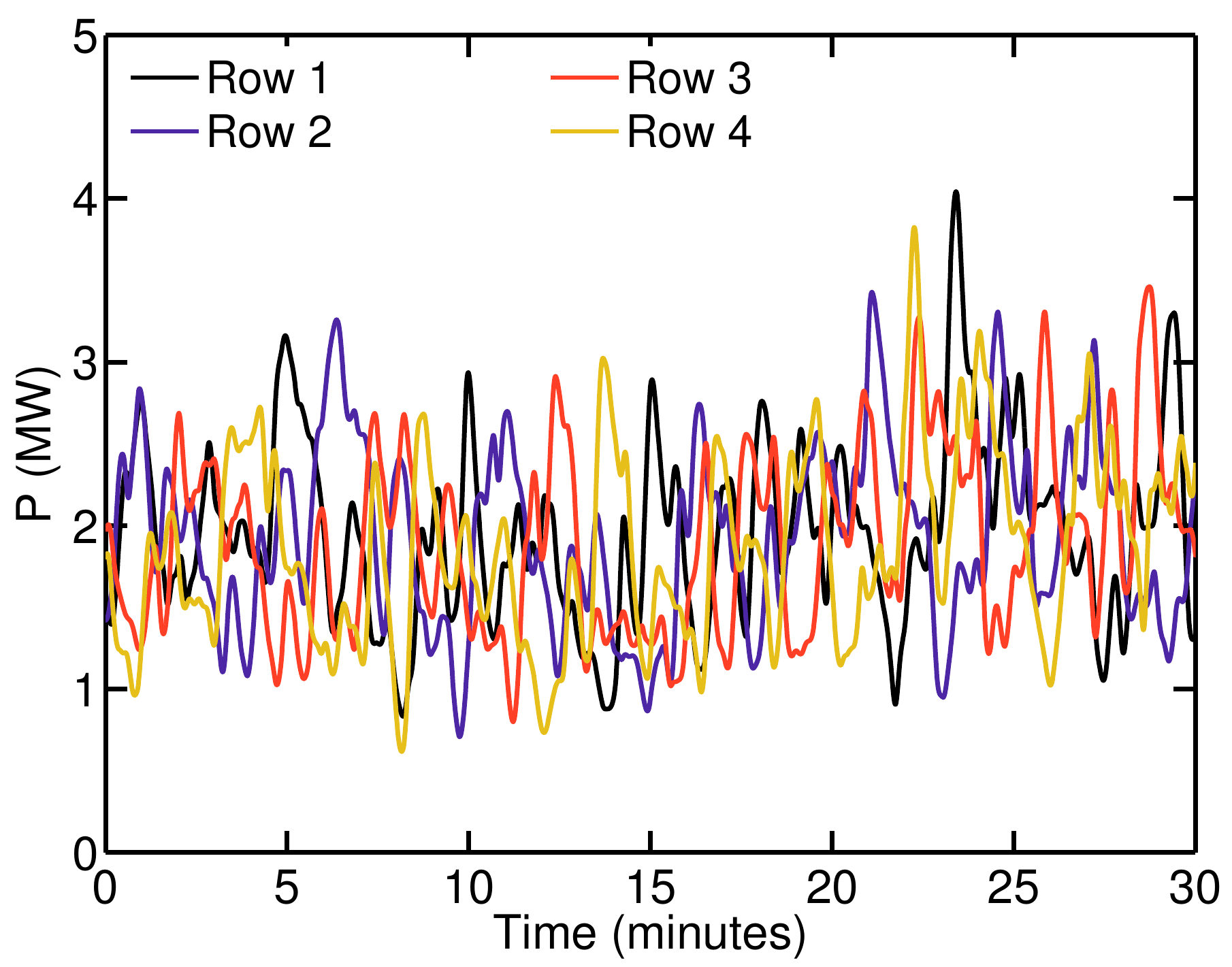}}
\caption{Power output of several turbines over time for simulation A2. The left and right column indicate the results for span-wise and stream-wise placed turbines. The upper and lower panels indicate a 5 minute moving average of the power output over a 6 hour period, and the instantaneous power output over a 30 minute interval, respectively.}
\label{figure3}
\end{figure*}

A typical snapshot of the flow is given in figure \ref{figure1} and reveals that the simulated flow is highly turbulent suggesting important time variations. Figure \ref{figure2}, which shows an instantaneous power signal of one of the turbines in simulation A2, confirms that the power output fluctuates strongly over a $6$ hour period. We note that the $0$ time at the beginning of the graph does not indicate the beginning of the simulation, but is already in the statistically stationary state. This has been done for all figures in which a start time of $t=0$ is indicated on the horizontal axis. This variability occurs even though in this simulations a fixed pressure gradient drives the flow. In real applications, such variability must be modulated further by daily cycle variations, large-scale meteorological phenomena such as fronts, etc. The variability we study is focused on the micro-scale fluctuations which, as can be readily seen, are significant by themselves. The top panels of figure \ref{figure3} shows a 5 minute moving-average of the power of the different turbines in that simulation and reveal that the power output of span-wise turbines seems to be rather uncorrelated, while there is a very strong correlation in the stream-wise direction. The lower panels in figure \ref{figure3} show the instantaneous power outputs over a short time domain and again show no correlation for span-wise placed turbine. In the stream-wise direction a maximum in the power output is observed in subsequent turbines with a certain time delay. This time delay is related to the travel time between the subsequent turbines and the large scale patterns formed in the ABL. 

Next, we compute frequency spectra of several signals $y(t)$, defined as usual according to
\begin{equation}
	\Phi \equiv | \hat{y} (\omega)|^2, ~~~ {\rm with ~~~~}
	\hat{y}(\omega) = \int_\infty^\infty y(t) \exp^{-i\omega t}dt,
\end{equation}
and evaluated using Fast Fourier Transform (FFT). The signals considered, $y(t)$ in most cases is the power output signal $P(t)$ of the turbines, but we also present some velocity spectra. To obtain better convergence of the spectra for the low frequencies, a Welch's averaged modified periodogram method in which the time signal is divided in 8 segments and windowed by a Hamming window, is used. Subsequently, we average the spectra of all the wind-turbines. They are statistically equivalent, since we use periodic boundary conditions in both horizontal directions. 

In order to determine the spectra for the entire wind-farm, or for aggregates over rows or columns, we first determine signals from the  turbine power output over time for the considered aggregates, by adding over the corresponding individual turbine signals and dividing by the number of turbines in the aggregate considered. Subsequently, the aggregate spectrum is calculated from these signals. In order to identify the different spectra we denote them by $\Phi_{X}^{Y}$. The lower index $X$ indicates that the spectrum is for a single turbine ($X=T$), a row of turbines ($X=R$), a column of turbines ($X=C$) or for all turbines together ($X=A$). The upper index $Y$ indicates the variable, i.e. whether we refer to the power output $Y=P$, the stream-wise velocity averaged over the turbine surface $Y=U_D$, the time averaged stream-wise velocity averaged over the turbine surface $U_D^T$, or the stream-wise velocity at the turbine center $U_C$ is used to calculate the spectrum, see also table \ref{table2} for an overview.

\begingroup
\squeezetable
\begin{table}
 \caption{The frequency spectra in this paper are identified using the notation $\Phi_{X}^Y$. The index $X$ indicates which turbine groups are considered in the determination of the spectrum, and the index $Y$ which signal is used. In the standard deviation $s_X$ the index $X$ has the same meaning.}
 \label{table2}
\begin{center}
\begin{tabular}{|c|c|l|}
 \hline
 Index &
 Symbol &
 Meaning \\
 \hline
 $X$ & $T$ 				& Single turbine \\
 $X$ & $R$ 				& Turbines in a row (span-wise flow direction) \\
 $X$ & $C$ 				& Turbines in a column (stream-wise flow direction) \\
 $X$ & $A$ 				& All turbines \\
 $Y$ & $P$ 				& Power output\\ 
 $Y$ & $U_D$ 				& Stream-wise velocity averaged over the turbine surface \\ 
 $Y$ & $U_D^T$ 			& Time averaged stream-wise velocity averaged over the turbine surface \\ 
 $Y$ & $U_C$ 				& Stream-wise velocity at turbine center \\ 
 \hline
\end{tabular}
\end{center}
\end{table}
\endgroup

\begin{figure}
\subfigure[]{\includegraphics[width=0.47\textwidth]{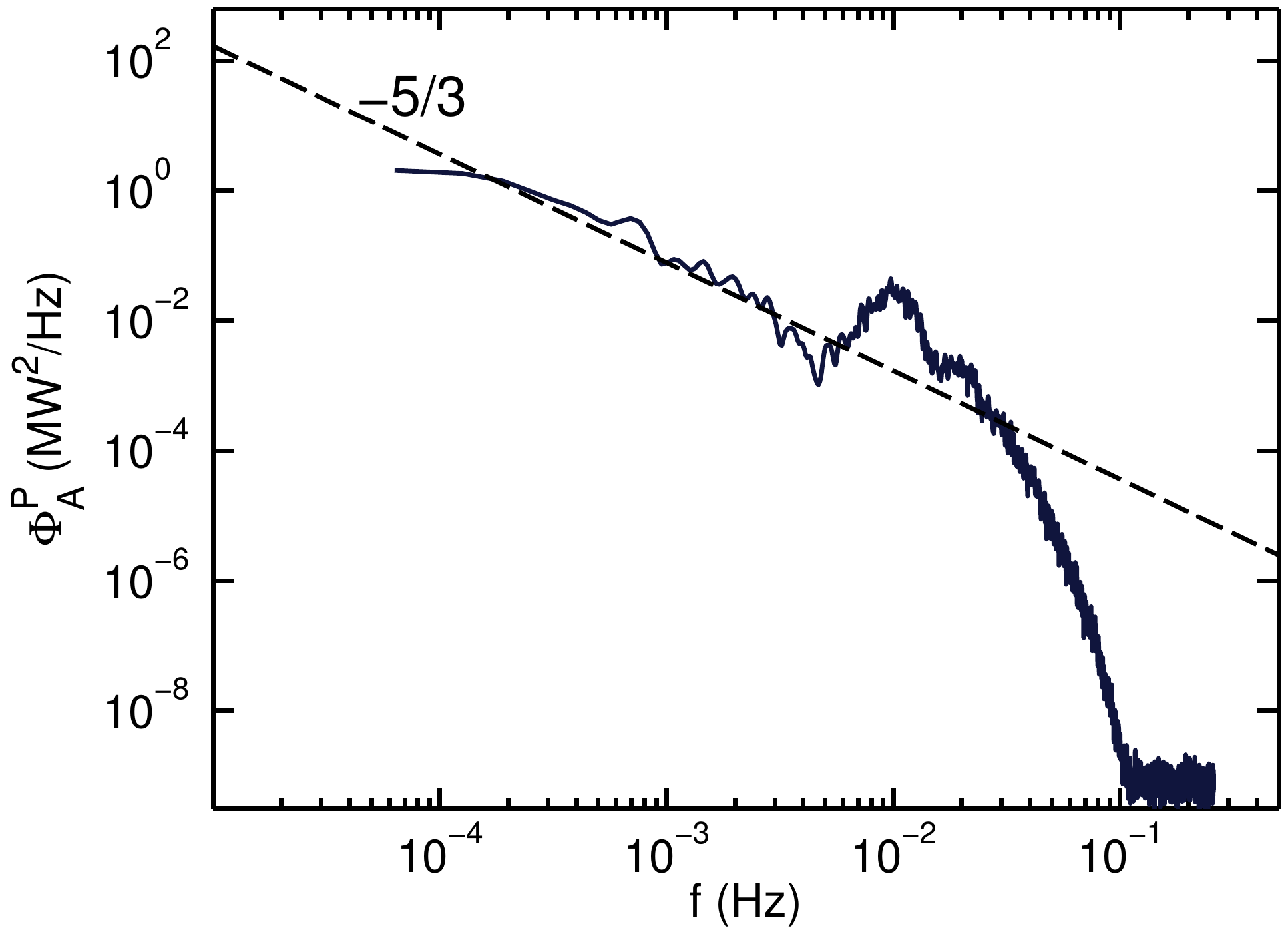}}
\subfigure[]{\includegraphics[width=0.47\textwidth]{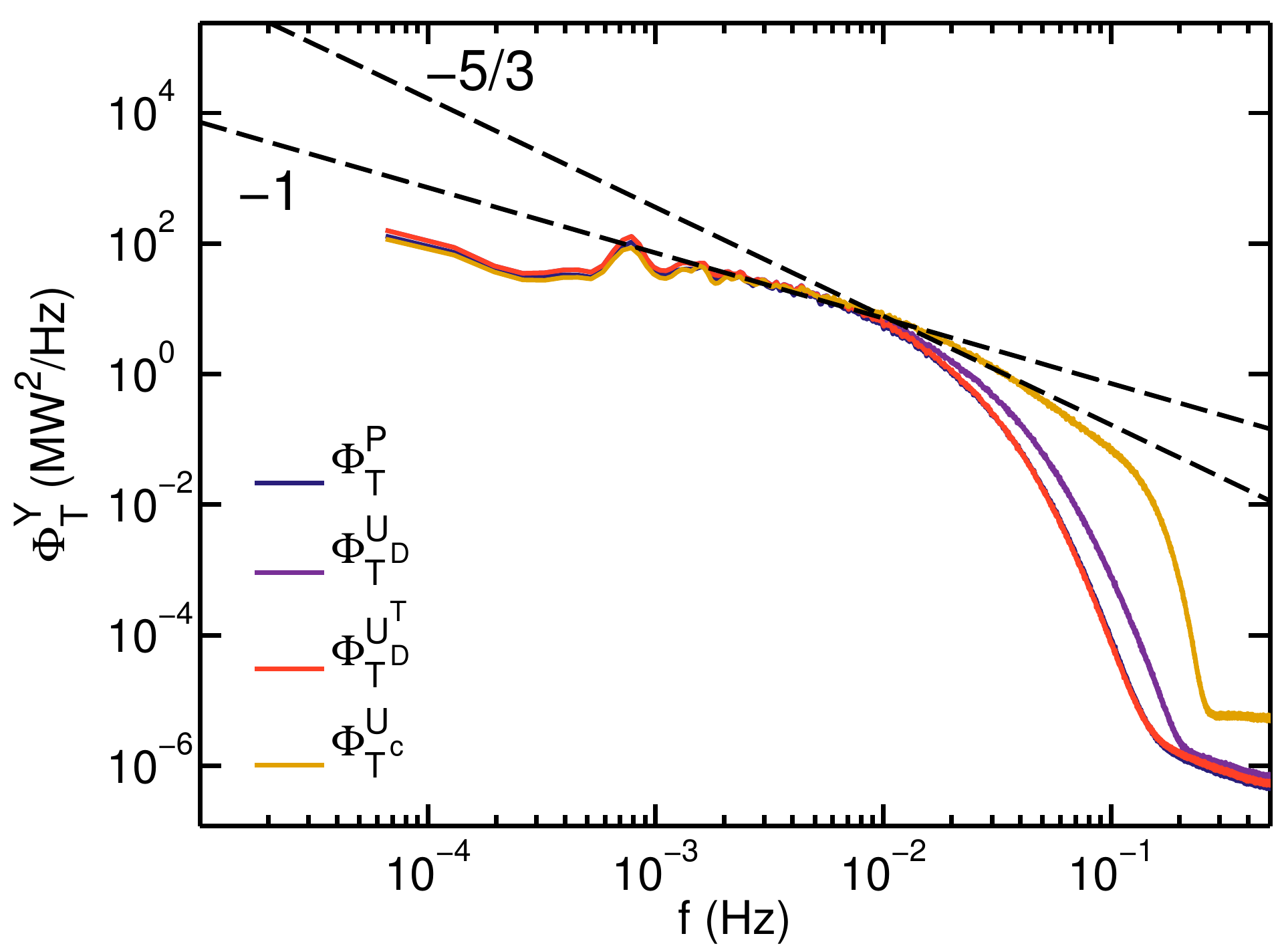}}
 \caption{(a) Frequency spectrum of the total power output of a very large wind-turbine park (simulation J1), showing an approximate $-5/3$ scaling for lower frequencies. (b) Frequency spectra of various signals from a single turbine from simulation E2. The figure shows the power output $\Phi_T^{P}$, the stream-wise velocity at turbine center $\Phi_T^{U_C}$, the stream-wise velocity averaged over the turbine surface $\Phi_T^{U_D}$, and the time averaged stream-wise velocity averaged over the turbine surface $\Phi_T^{U_D^T}$. At around $f \sim 0.006$ (Hz) the spectrum for the stream-wise velocity at the turbine center transitions from a steep slope (including a very short $-5/3$ scaling near $f \sim 0.04$ (Hz)) to a $-1$ scaling. Note that $\Phi_T^{U_C}$, $\Phi_T^{U_D}$, and $\Phi_T^{U_D^T}$ are arbitrarily shifted to allow comparison of the shape of the $\Phi_T^{P}$ spectrum.}
\label{figure4}
\end{figure}

Figure \ref{figure4}a shows the power spectrum for the aggregate consisting of all wind turbines, i.e. from the overall average over all wind-turbines in a very large wind-turbine farm, case J1. Remarkably, the spectrum shows a power-law behavior over a range of frequencies between $4\times10^{-3}$ and $2\times10^{-4}$ (Hz) with a slope of (approximately) $-5/3$. Such `Kolmogorov scaling' is consistent with similar observations made in field-scale operational wind-parks made by Apt \cite{apt07}. Another interesting feature is the peak in the spectrum. As we do not model the rotation of the blades we can exclude that this peak is due to the helicoidal tip vortices that are shed by the blades. These characteristics differ from properties of the velocity spectra in the turbine wake \cite{cha10,chu12b} that have revealed a transition from a $-5/3$ Kolmogorov spectrum towards the $-1$ range at lower frequencies. Below we will document such a transition when focusing on velocity at a single wind-turbine from the present LES. 

In the remainder of the paper we will further analyze the spectral behavior presented in figure \ref{figure4}a by considering various aggregates of wind-turbines within the wind-farm. 
To assess resolution issues as well as to provide a reference case, we first discuss the spectra for a single wind-turbine in the array, see section \ref{section4a1}, before returning to the spectrum of the entire wind-farm, see section \ref{section4a2}, where we give special attention to the relation between the power output in the rows and columns of turbines and the difference between a staggered and an aligned wind-park. Subsequently, we will discuss the effect of varying wind-speed (section \ref{section4a}) and the effect of varying wind-direction (section \ref{section_new2} and \ref{section_new3}), before comparing with results for a finite size wind-farm in section \ref{section4b}). 
We first discuss the spectra for a single wind-turbine in the array, see section \ref{section4a1}, before returning to the spectrum of the entire wind.

\subsection{Spectra from a single wind-turbine in the array} 
\label{section4a1}

\begin{figure}
\subfigure[]{\includegraphics[width=0.47\textwidth]{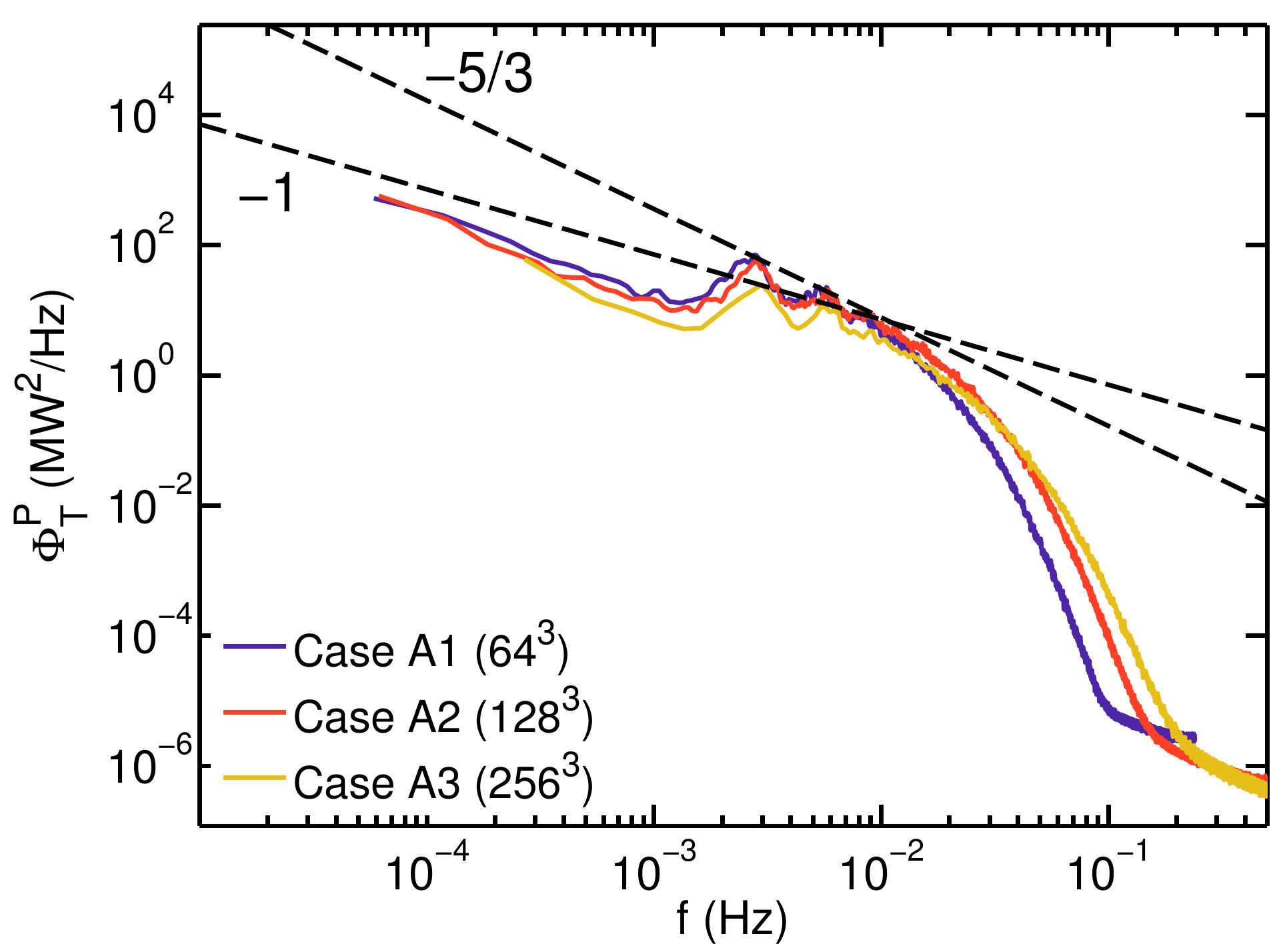}}
\subfigure[]{\includegraphics[width=0.47\textwidth]{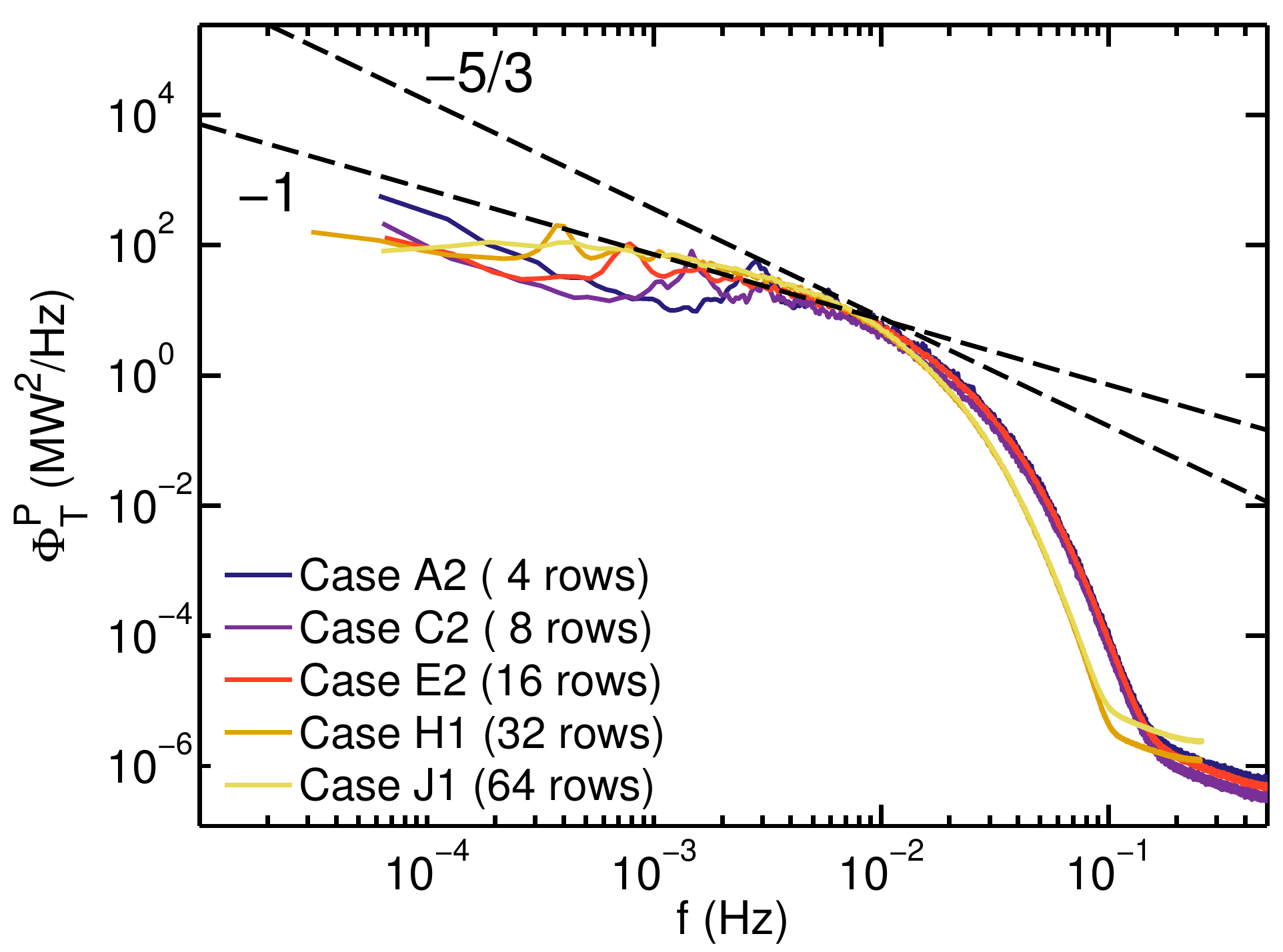}}
 \caption{a) Frequency spectra of single turbine power signals using various spatial grid resolutions for the LES. The figure shows that the spectra obtained on different grid resolutions, i.e.\ the $A*$ simulations, reveal the same main features. As expected, the higher resolution simulations capture more of the higher frequencies. b) Comparison of the spectra for the power output per turbine obtained from simulations on different domain sizes, ranging from a domain with only 4 turbines in the stream-wise direction (simulation A2) to domains with 64 turbines in the stream-wise direction (simulation J1).}
\label{figure5}
\end{figure}

Figure \ref{figure4}b shows that the spectrum for the velocity in the center of the turbine follows a $-5/3$ scaling for about a factor of two, before it transitions towards a $-1$ scaling at lower frequencies. The figure reveals a very sharp drop in the spectrum near $f \sim 0.2$ (Hz), which is determined by the spatial resolution of the LES (for this case a grid-spacing of about $\Delta_x=23$m with a convection velocity of $U_D^T \sim 5.6$ m/s yields a time-scale of $\Delta_x / U_s \sim 4.1$ s). The higher frequencies are less pronounced in the spectrum of the power output of the turbine as it is determined from the disk averaged velocity over the turbine area including small time averaging. Figure \ref{figure5}a reveals that the main characteristics of the spectrum do not change when the resolution is increased, although one can see that some of the higher frequency components are better resolved in the higher resolution simulations. 

Next, we consider the effect of the domain size. Figure \ref{figure5}b shows that when the stream-wise length of the domain is increased the transition from the $-5/3$ spectrum towards the $-1$ scaling regime is captured better as the drop in the spectrum, which is related to the flow through time of the domain, shifts towards lower frequencies when the domain length is increased. The comparison reveals that the lower frequency components are captured better by simulations performed on a longer domain and that the drop in the spectrum in the low frequency part is due to the limited domain size in the stream-wise direction. The figure reveals that the lowest frequency peak created due to the limited domain length shifts by a factor of about $2$ and weakens when the domain length is doubled. These low frequency peaks are no longer visible for the spectrum obtained in the longest domain. Hence, most results presented in the remainder of the paper will be for longer domains of at least $L_x=4\pi$ with columns of 16 turbines in the computational domain (cases E2 and  F2).  

\subsection{Spectra in various aggregates} \label{section4a2}

\begin{figure}
\subfigure[]{\includegraphics[width=0.47\textwidth]{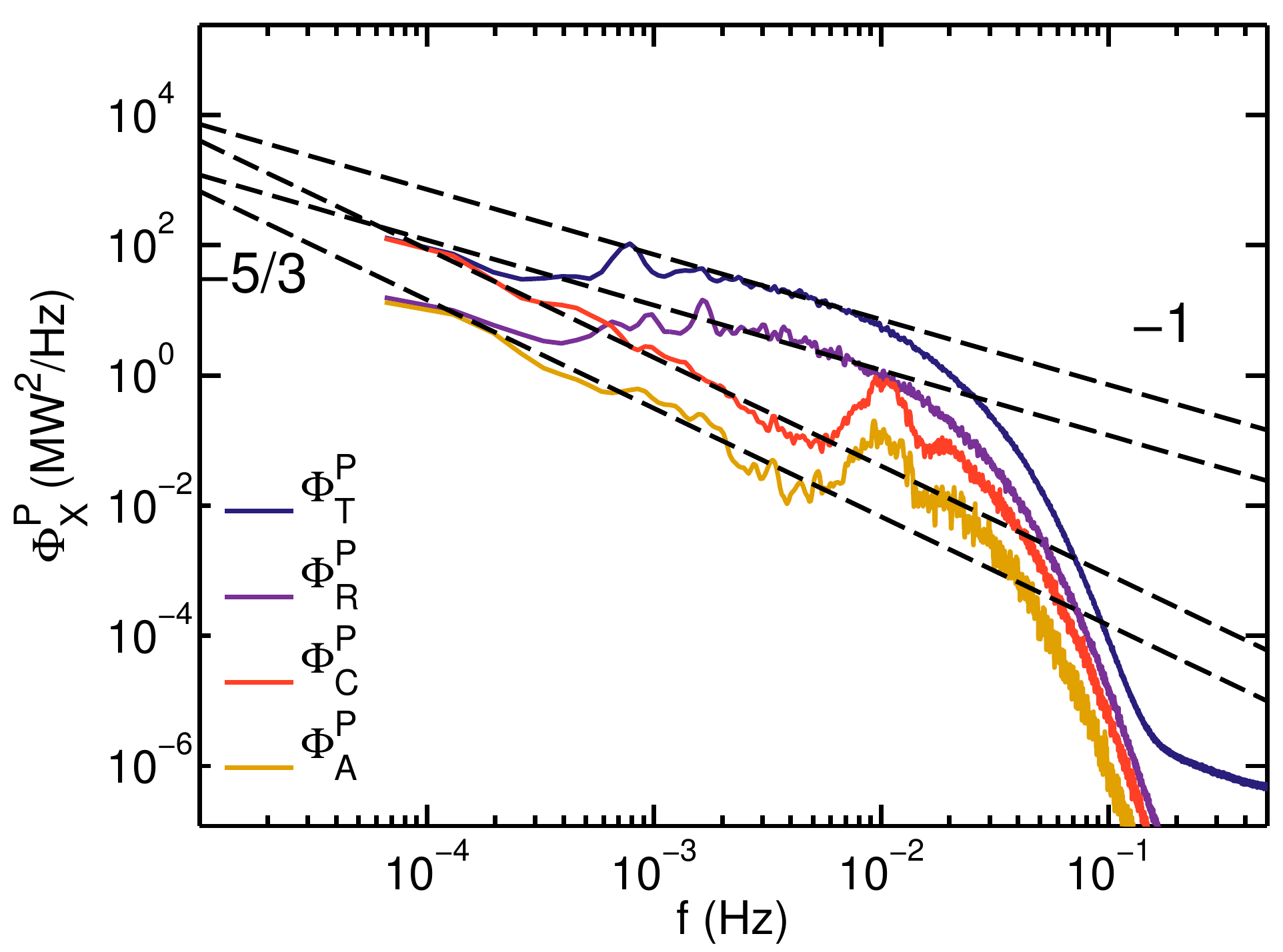}}
\subfigure[]{\includegraphics[width=0.47\textwidth]{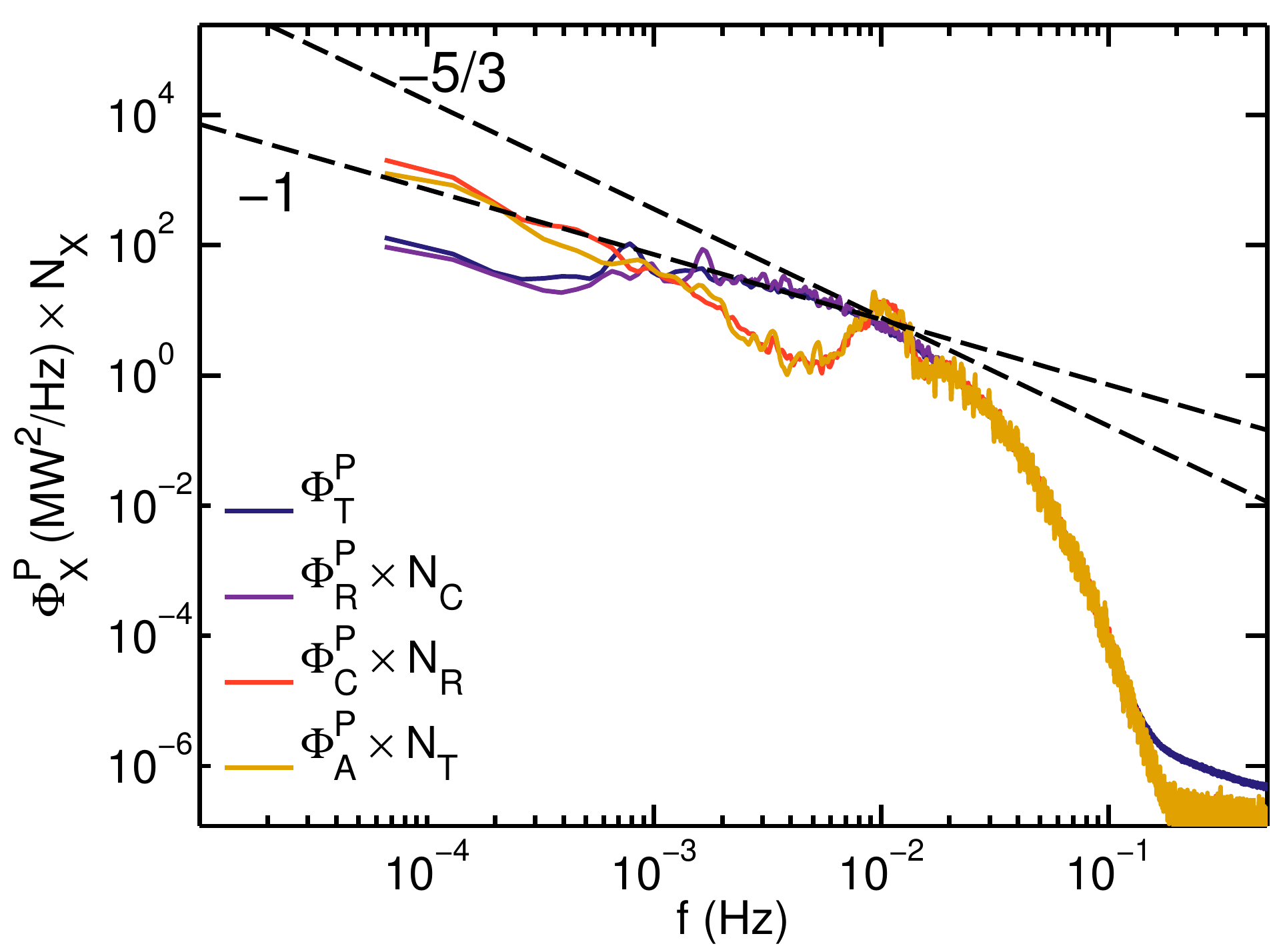}}
 \caption{a) Comparison of the spectra of the power for a single turbine $\Phi_T^P$, a row of turbines $\Phi_R^P$ (in the span-wise flow direction), a column of turbines $\Phi_C^P$ (in the stream-wise flow direction), and the entire wind-farm $\Phi_A^P$, for case E2. b) Same as panel a, but now the spectra for the row, column, and total output have been multiplied by $N_C$, $N_R$, and $N_T$, respectively, which is the expected decrease of the spectra if the power output of all turbines were uncorrelated.}
\label{figure6}
\end{figure}

In this section we consider spectra of power considering various aggregates. Mean power and variance of power fluctuations in all cases have already been
presented in table \ref{table1} and are discussed further in section \ref{statistics}.  Figure \ref{figure6}a compares the spectrum of the power output fluctuations of a single turbine with the spectrum of the power output of all turbines in a row (turbines in the span-wise flow direction), a column (turbines in the stream-wise flow direction), and in the entire wind-park. As is expected when adding a number of signals, there is a significant reduction in the power fluctuations when aggregates of increasing number of turbines are considered. 

\begin{figure}
\subfigure{\includegraphics[width=0.47\textwidth]{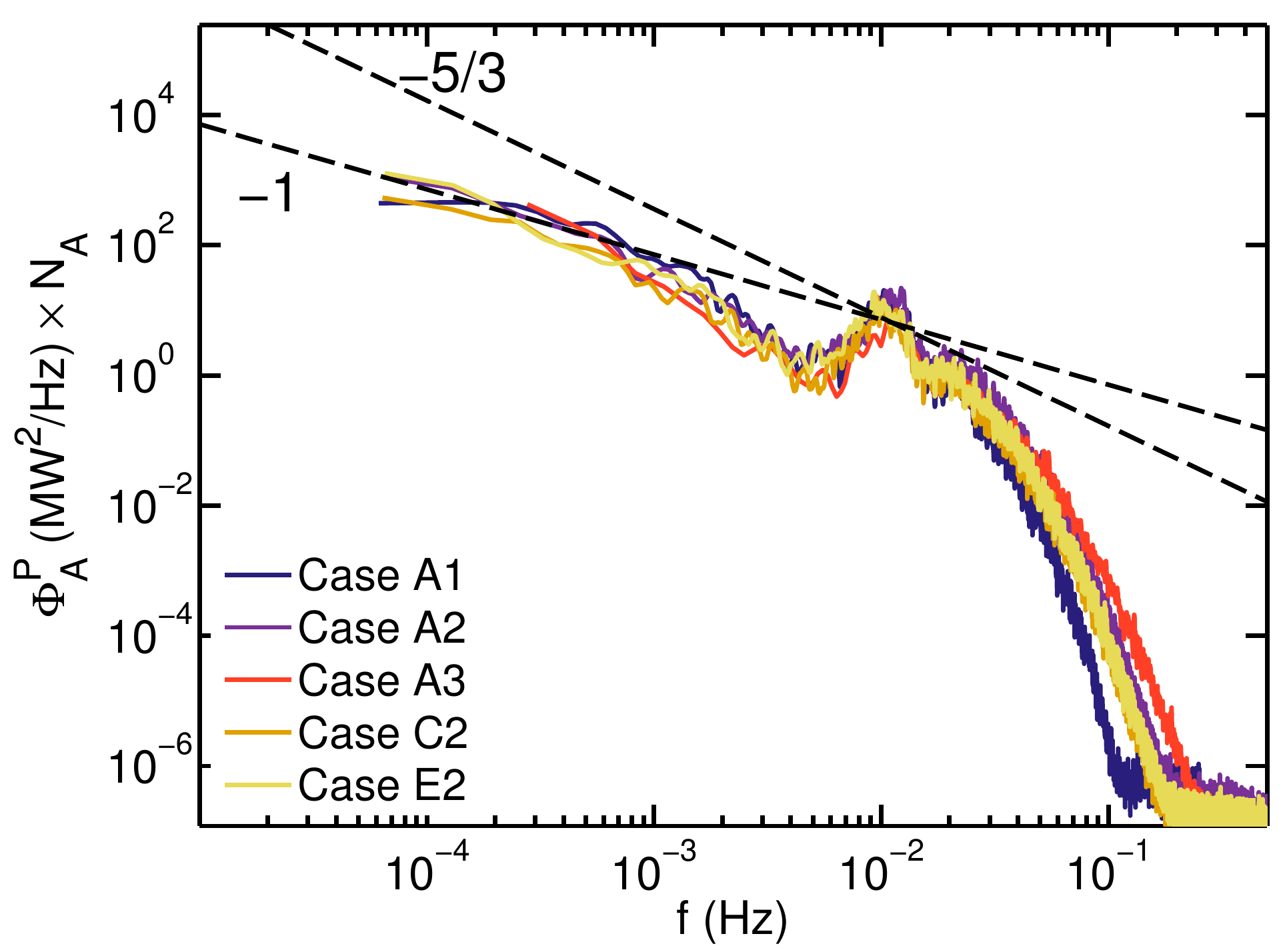}}
 \caption{Power spectra obtained for the entire wind-park, from simulations with different domain sizes and different spatial resolutions. Note that all spectra have been multiplied with the respective number of turbines.}
\label{figure7}
\end{figure}

We attempt to disentangle the behavior of the spectra by a rescaling that should be exact if the power signals among the turbines were statistically uncorrelated. In such a case, one should be able to relate all by multiplying the spectra by the number of turbines $N_X$ involved in the respective averaging operation, i.e.\ the total spectrum for the uncorrelated case would be given by
\begin{equation}
\Phi_{X}^P = N_X \times \Phi_T^P
\end{equation}
where $X$ indicates that a row ($R$), column ($C$), or the entire wind-farm ($A$) is considered. The results of such rescaled spectra are shown in figure \ref{figure6}b. The rescaled spectrum of the power output of a row of turbines looks almost identical to the spectrum of a single wind-turbine, i.e.\ $\Phi_{R}^P \approx N_R \Phi_{T}^P$. Consistent with figure \ref{figure3} which shows that there seems to be no correlation between the power output of the different turbines placed in the span-wise direction, the rescaled power spectrum for the rows of turbines agrees well with that of a single turbine. We further confirmed this observation by calculating the cross-correlation between the power output of the different span-wise turbines, which indeed revealed essentially no correlation. 

Next, we consider the aggregate spectrum of turbines placed in the stream-wise direction. These results show an unexpected strong reduction of the spectrum for frequencies below the 'peak' in the spectrum, for $f < 7\times 10^{-3}$ Hz. This feature is then also observed in the total power output of the wind-farm because the power output of the different rows is uncorrelated. To verify that this strong reduction of the intermediate frequencies is not related to the length of the domain size, or the numerical resolution, we compared the spectrum obtained for the entire wind-farm for different cases in figure \ref{figure7}. This figure confirms that the reduction is unrelated to the used domain size and numerical resolution. 

\begin{figure}
\subfigure[]{\includegraphics[width=0.47\textwidth]{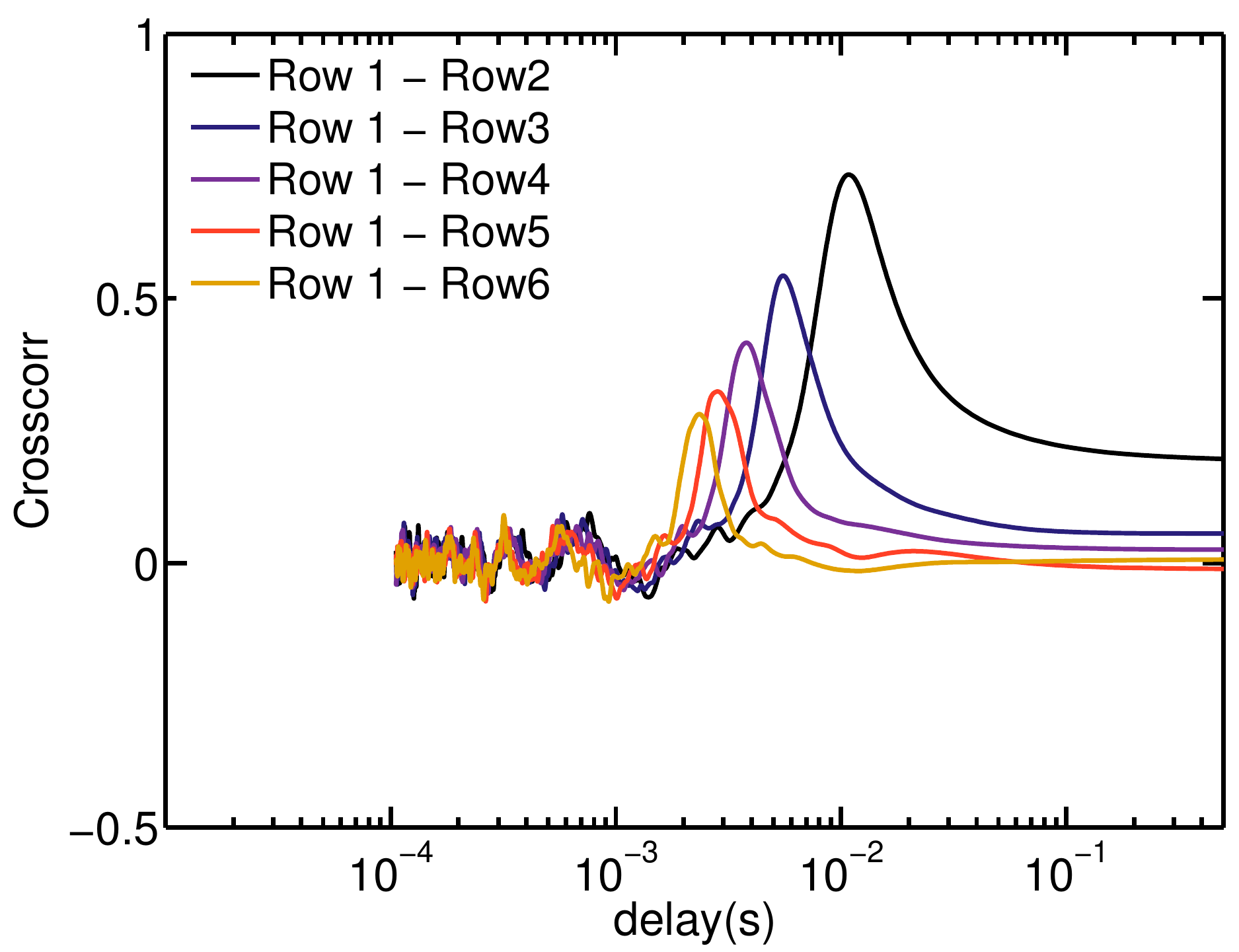}}
\subfigure[]{\includegraphics[width=0.47\textwidth]{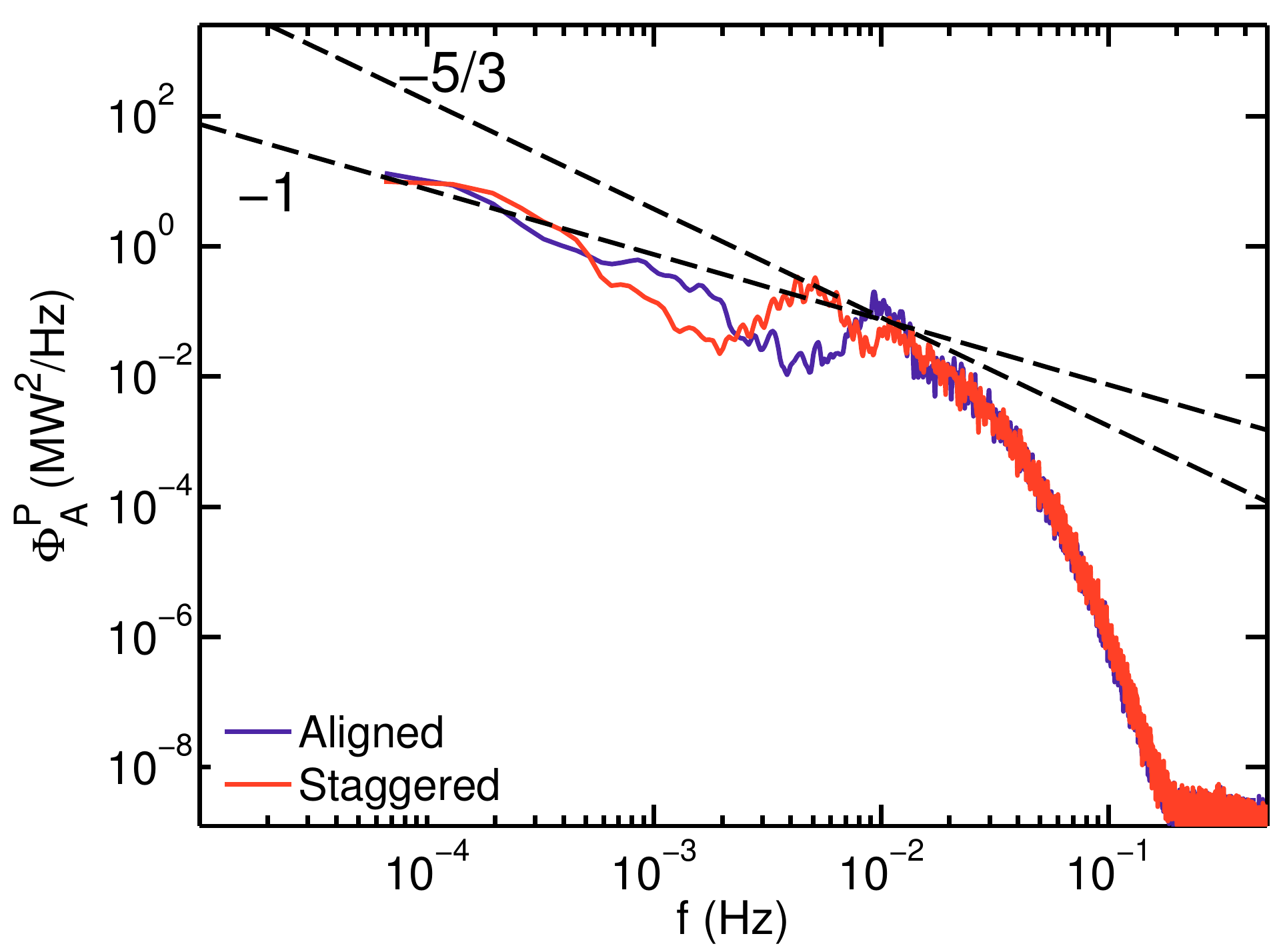}}
 \caption{a) The cross-correlation between the power output of subsequent down-stream wind-turbines in simulation E2. The figure shows that there is a significant peak correlations between the turbine power output of turbines placed closest to each other in the stream-wise direction at a particular frequency that decreases for the more distant wind-turbine pairs. b) Comparison of the spectra of the power output of the entire farm for an aligned (E2) and a staggered (F2) wind-farm shows that the distance between subsequent downstream wind-turbines influences the output spectrum.}
\label{figure8}
\end{figure}

The position of the peak and the subsequent drop in the spectrum seem to be related to the travel time between the wind-turbines of large scale structures present in the ABL. To confirm this hypothesis we calculated the cross-correlation between the power output of consecutive turbines in the down-stream direction, see figure \ref{figure8}a. The peaks in this figure can be seen as the delay for high or low velocity wind-patches (large scale flow structures) to travel from one turbine row to the next. Figure \ref{figure8}a reveals that the correlation is strongest between two consecutive turbines and is significant up to several downstream turbines. The drop in the aggregate spectrum is only visible when an array of turbines in the stream-wise direction is considered. In that case the lower frequencies in the aggregate spectrum become weaker because similar power signals, with different delay times, are added together and each time delayed signal decreases the energy in its corresponding frequency band of the aggregate spectrum. When this procedure is performed with identical signals the drop of the spectrum increases with the number of turbine rows. When the power output of the turbines obtained from the LES are considered this does not occur because the power output of turbines far downstream becomes uncorrelated, see figure \ref{figure8}a. Interestingly, a spectrum close to a $-5/3$ scaling in the aggregate spectrum arises for the lower frequency range, after these averaging operations are carried out, due to these complex interactions. 

Figure \ref{figure8}b compares the power output of a simulation in which the turbines are aligned, i.e.\ placed directly in each others wake, and a staggered alignment, in which the turbines in each subsequent downstream row are shifted by half a turbine spacing. Both spectra show an approximate $-5/3$ scaling in the low frequency range after a 'peak' in the spectrum, which is shifted by a factor of $2$ when the turbines are placed in a staggered arrangement. The shift of the peak by a factor of $2$ confirms that the dominant spectral peak is related to the downstream distance and travel time between the turbines. 

\subsection{Mean power and variance}
\label{statistics}
\begin{figure}
\subfigure[]{\includegraphics[width=0.47\textwidth]{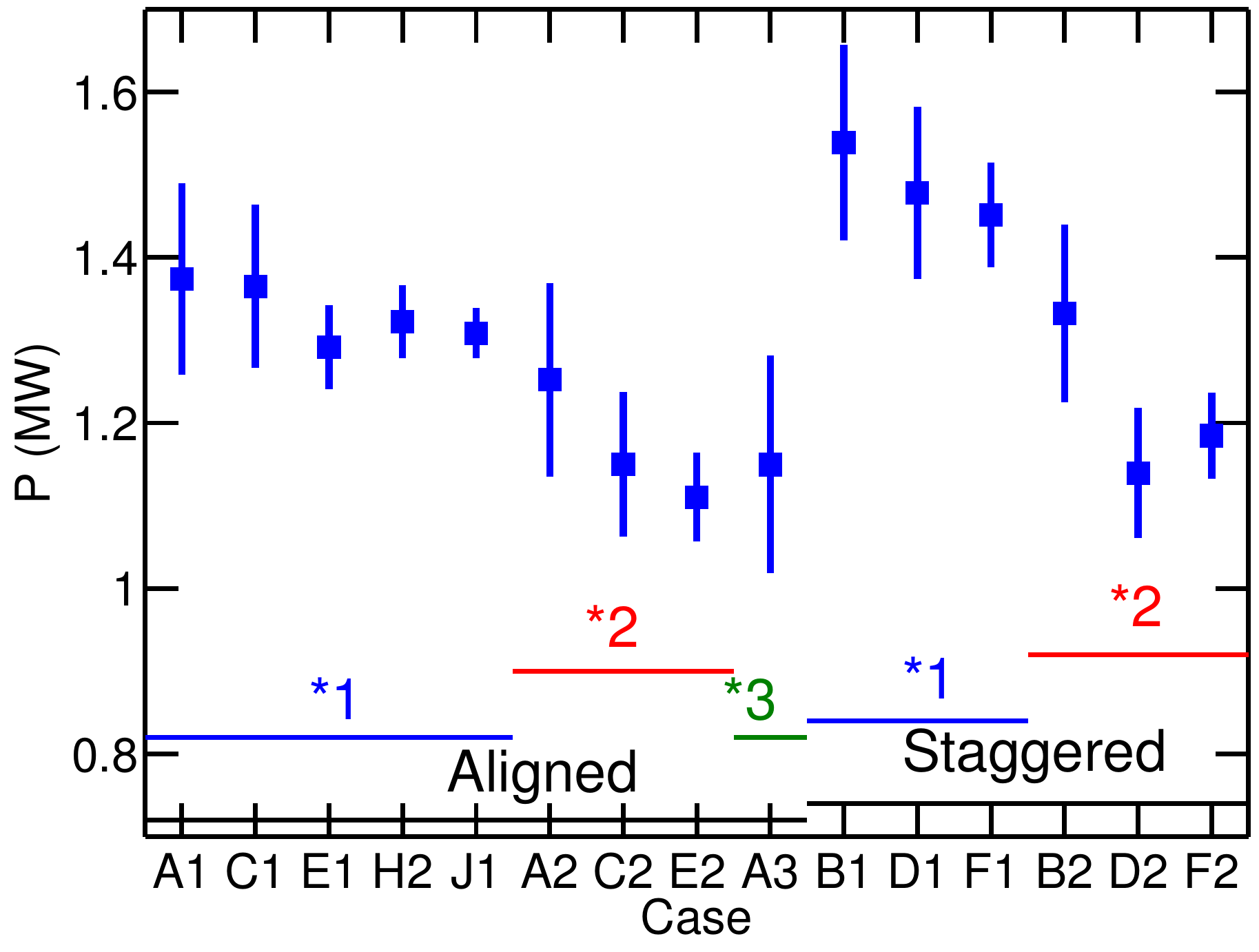}}
\subfigure[]{\includegraphics[width=0.47\textwidth]{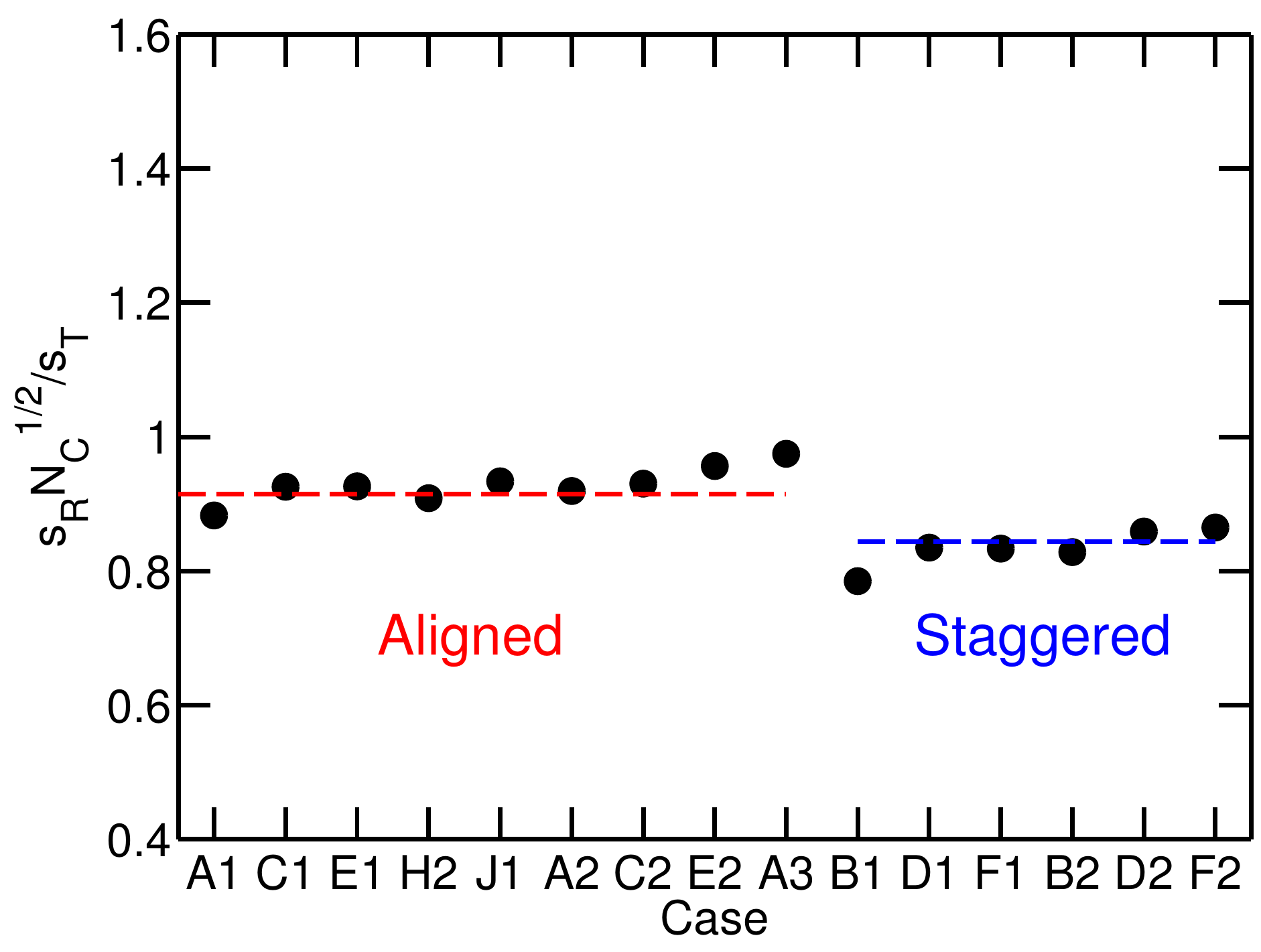}}
\subfigure[]{\includegraphics[width=0.47\textwidth]{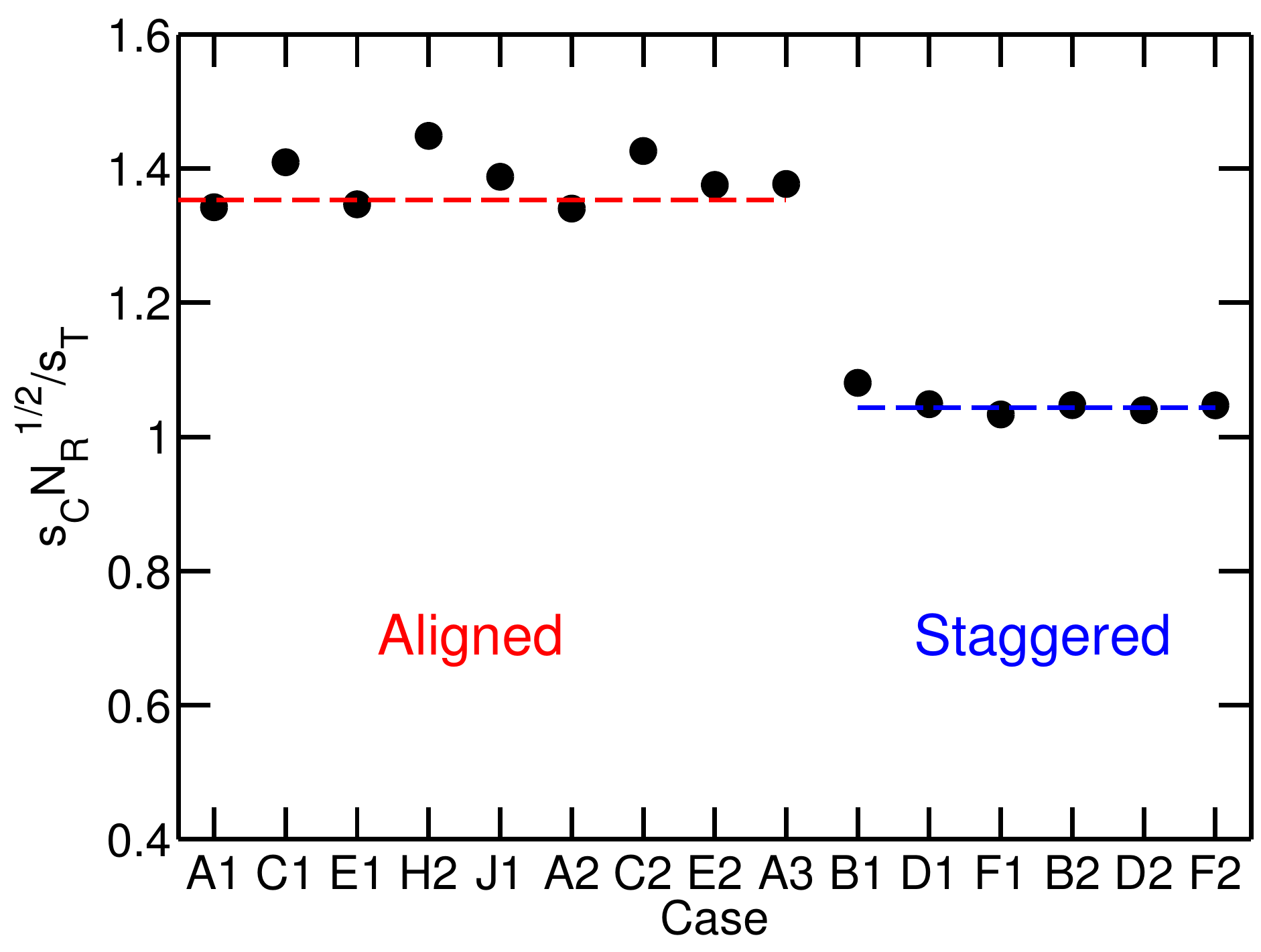}}
\subfigure[]{\includegraphics[width=0.47\textwidth]{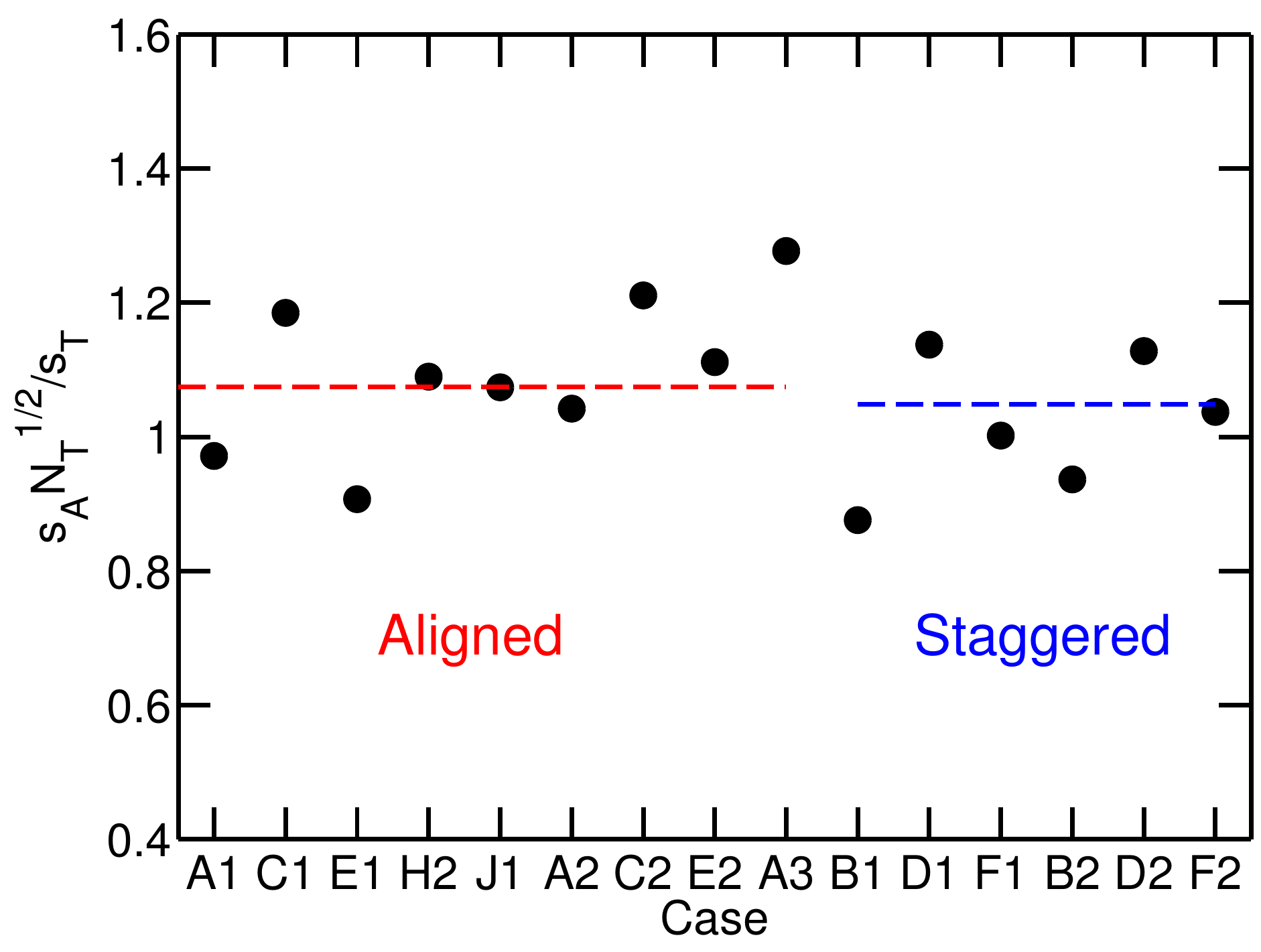}}
 \caption{(a) The average power output of the turbines in the infinite wind-farm simulations. The indicated uncertainty is the standard deviation in the power output of all turbines. The standard deviation of the power output for a row of turbines (b), a column of turbines (c) and all turbines (d) compared with the the standard deviation of the power output of a single turbine corrected for the number of turbines, see eq.\ (\ref{s_relation}). In each panel the cases from left to right are for aligned and staggered configuration with increasing length and resolution in both categories, see also table \ref{table1}.}
\label{figure9}
\end{figure}

Table \ref{table1} presents overall statistics on the power signals in the simulations, which are also shown in figure \ref{figure9}. Figure \ref{figure9}a reveals that the time-average turbine power output $\langle P \rangle$ slightly decreases with increasing resolution and domain length and is reasonably well converged for the standard cases (cases E2 and  F2) we are considering. In addition, the figure reveals that the average turbine power output is slightly higher in an staggered wind-farm than in an aligned wind-farm, in agreement with Refs. \cite{mey10,yan12,wu13,ste14b}.  Also, we find a very good agreement between the mean power output for the infinite (table \ref{table1}) and the finite size wind-farm (presented later in table \ref{table3}). For the finite size wind-farm we obtain an average power output in row 4 to 12, which are closest to the fully developed state, of $1.186$ MW compared to $1.150$ MW for the infinite wind-farm (case C2). For the finite size wind-farm we can compare this to the power output of the first row ($1.94$ MW) to find that turbines in the fully developed regime produce roughly $61\%$ of the power produced by turbines on the first row.
 
Table \ref{table1} also shows the standard deviation $s_X$ of the power of several aggregates, calculated as
\begin{equation}
s_X= \langle (P_X(t) - \langle P_X \rangle )^2 \rangle^{\frac{1}{2}},
\end{equation}
where the index $X$ has the same meaning as indicated before, see also table \ref{table2}, and $\langle .. \rangle$ indicates time-averaging over the statistical stationary part of our data as described in section \ref{section4a}. One expects that the standard deviation of an aggregate of wind-turbines is related to the standard deviation of a single wind-turbine as 
\begin{equation}\label{s_relation}
s_X= \frac{s_T}{\sqrt{N_X}}
\end{equation}
Figure \ref{figure9}b reveals that the standard deviation of the power output of a row of turbines decrease slightly more than expected for uncorrelated turbines. We think this is related to the streaks that are formed in an ABL, which means that some turbine feel a higher incoming velocity than other turbines. A close look at figure \ref{figure6}b shows that the spectra for a row of turbines is slightly below the spectrum for a single turbine in the low frequency range, which is consistent with this view. Figure \ref{figure9}c shows that the reduction of the standard deviation is less than expected for turbines placed downstream of each-other, which is due to the correlation of the power output of turbines in the downstream direction, see figure \ref{figure8}a. Because the downstream distance between the turbines is larger in a staggered wind-farm than in an aligned wind-farm the effect is more pronounced in aligned wind-farms. The uncertainty in the standard deviation for the entire wind-farm is larger, which is indicated by the scatter of the data in figure \ref{figure9}d, because the result cannot be averaged over different rows or columns. The data in this panel do not reveal a clear trend of the normalized standard deviation for the total wind-farm power output compared to the power fluctuations of a single turbine. 

\section{Effects of external conditions}
In this section we consider the effects of various external conditions on the main observations presented before. 
In particular we examine the robustness of the $-5/3$ range observed at low frequencies for aggregate wind power fluctuations. 
We first discuss the effect of a varying wind-speed in section \ref{section4a}. Then we discuss the influence of the wind-direction in section \ref{section_new2} and of time-varying direction in section \ref{section_new3}. This will be followed by a short comparison with a finite size wind-park in section \ref{section4b}. 

\subsection{Time-varying overall wind-speed} \label{section4a}
\begin{figure}
\subfigure[]{\includegraphics[width=0.47\textwidth]{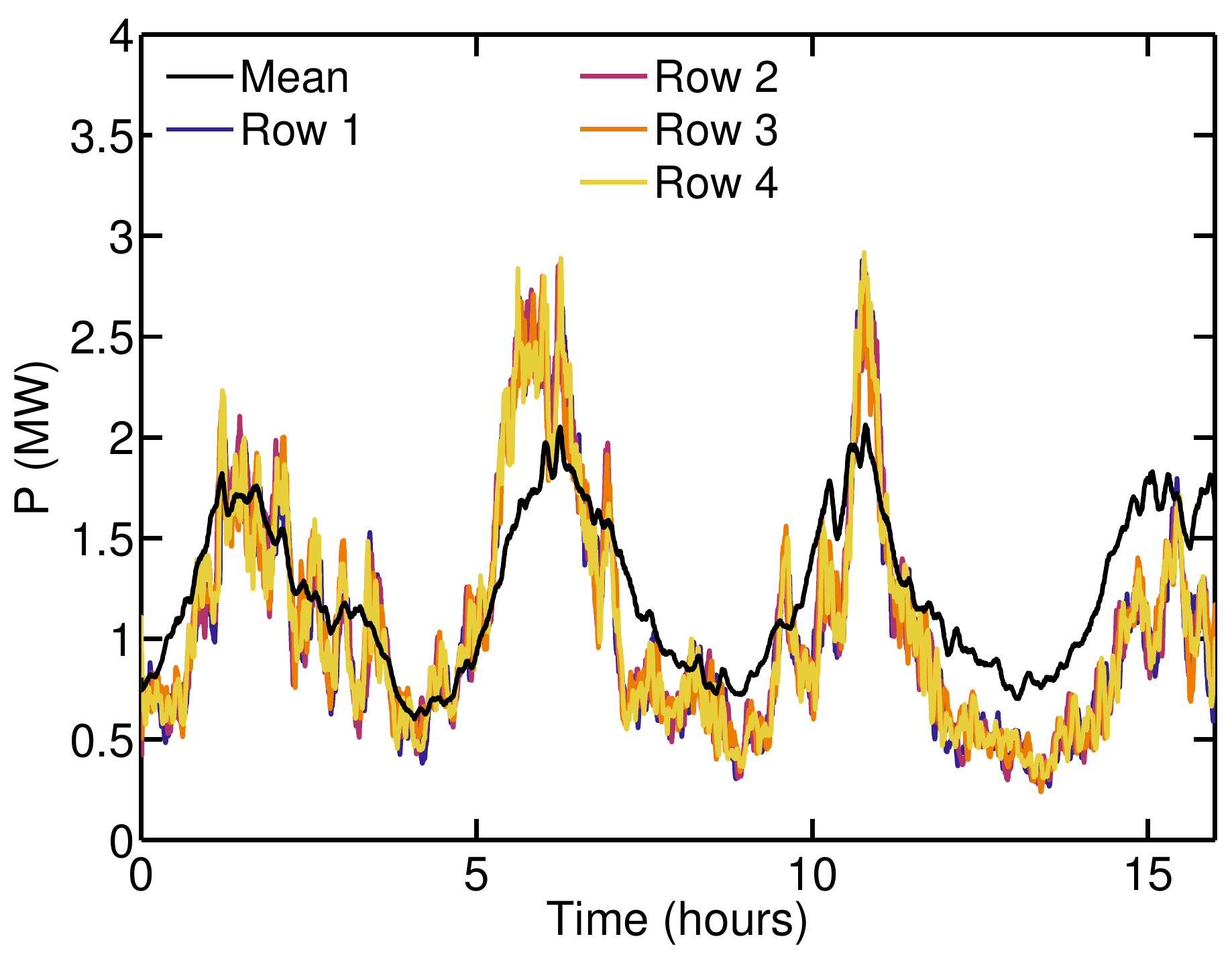}}
\subfigure[]{\includegraphics[width=0.47\textwidth]{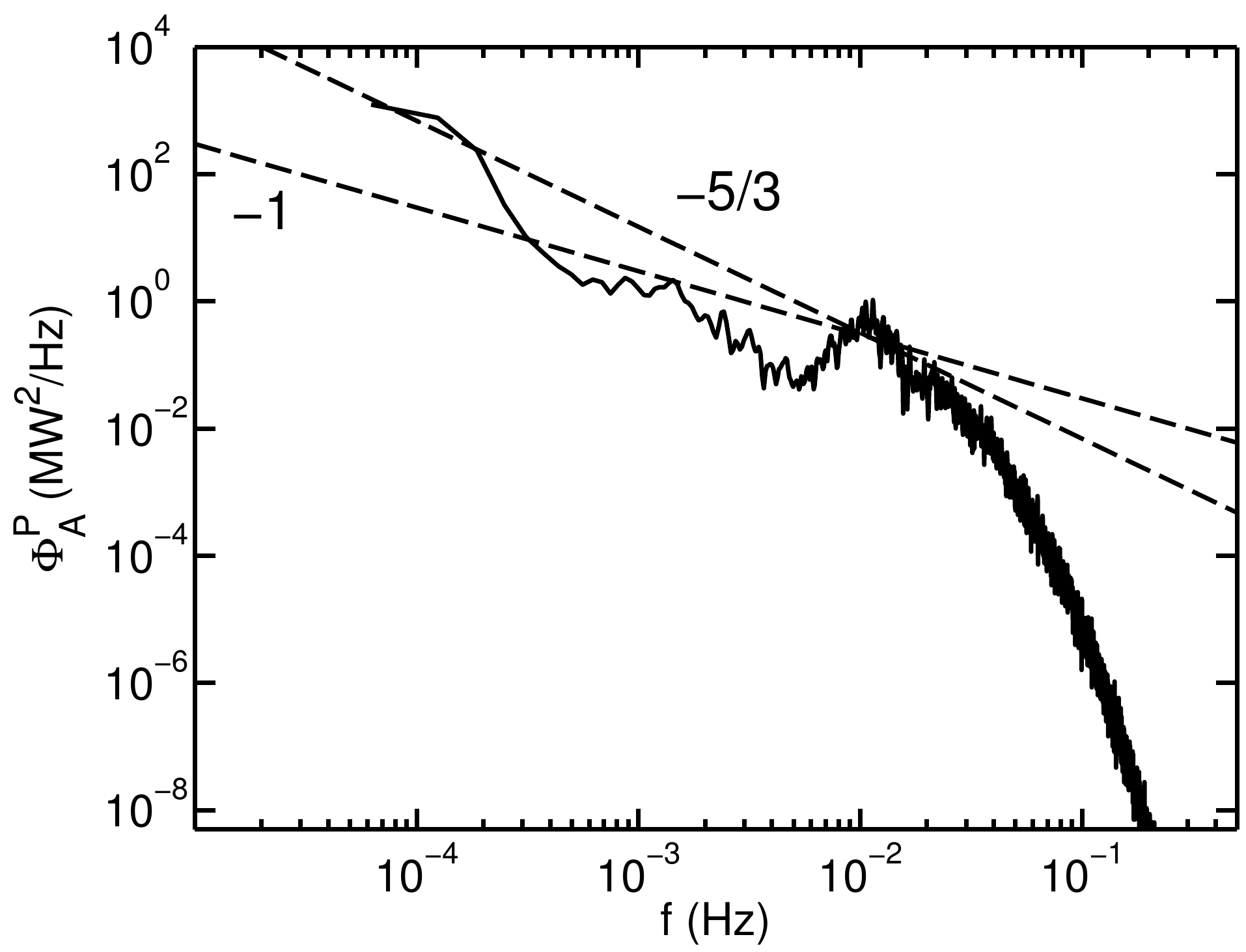}}
 \caption{(a) Power output of a selected number of turbines in the same column (different rows in stream-wise direction) and mean power output using a 5 minute moving averaging, over a 16 hour time period and reveals the slow time variation that is imposed on the mean wind-velocity. (b) Power spectrum $\Phi_A^P$ of the total wind-farm power output fluctuations. It is very similar to the one obtained in figure \ref{figure6}b.}
\label{figure10}
\end{figure}

In order to determine the influence of the wind-speed on the aggregate spectrum of the wind-farm we performed a simulation in which we modify the pressure forcing over time by setting $A=0.8$ in eq.\ \ref{eq_forcing1}. This results in a slowly changing average wind-speed in our domain. Figure \ref{figure10}a shows the average power output of the wind-farm, and of some selected wind-turbines, as function of time. In agreement with expectations, the figure shows that the average power slowly changes over time. From the figure we can see that average power changes from roughly $1$ MW to $2$ MW. This means that the average wind-speed changes with roughly $25\%$. Note that this change is significantly less than the change in the pressure forcing that is applied because the flow only changes slowly to the applied changes in the pressure gradient and this leads to a time lag and damping. The spectrum of the entire wind-farm is shown in figure \ref{figure10}b and reveals that the spectral peak is still present even though the travel time between the turbines now varies due to the changing average wind-speed. Our interpretation is that the periods with the strongest wind-speeds dominate the resulting spectrum, because higher wind-speed periods have significantly higher power output and power fluctuations than the low wind-speed periods, assuming that the turbines remain operating in regime II.

\subsection{Imposed wind-direction} \label{section_new2}
\begin{figure}
\subfigure[$\phi=0$ degrees]{\includegraphics[width=0.32\textwidth]{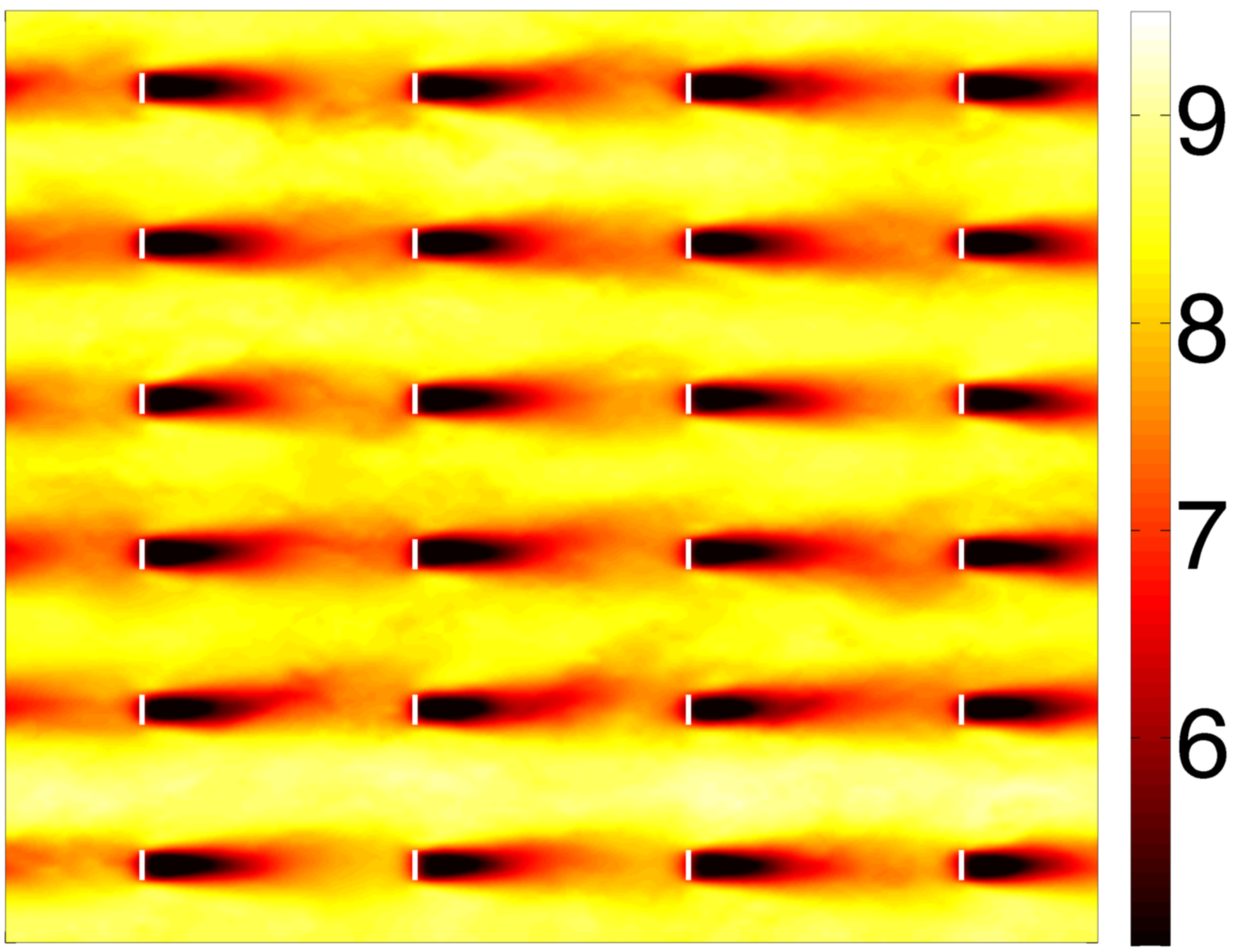}}
\subfigure[$\phi=50$ degrees]{\includegraphics[width=0.32\textwidth]{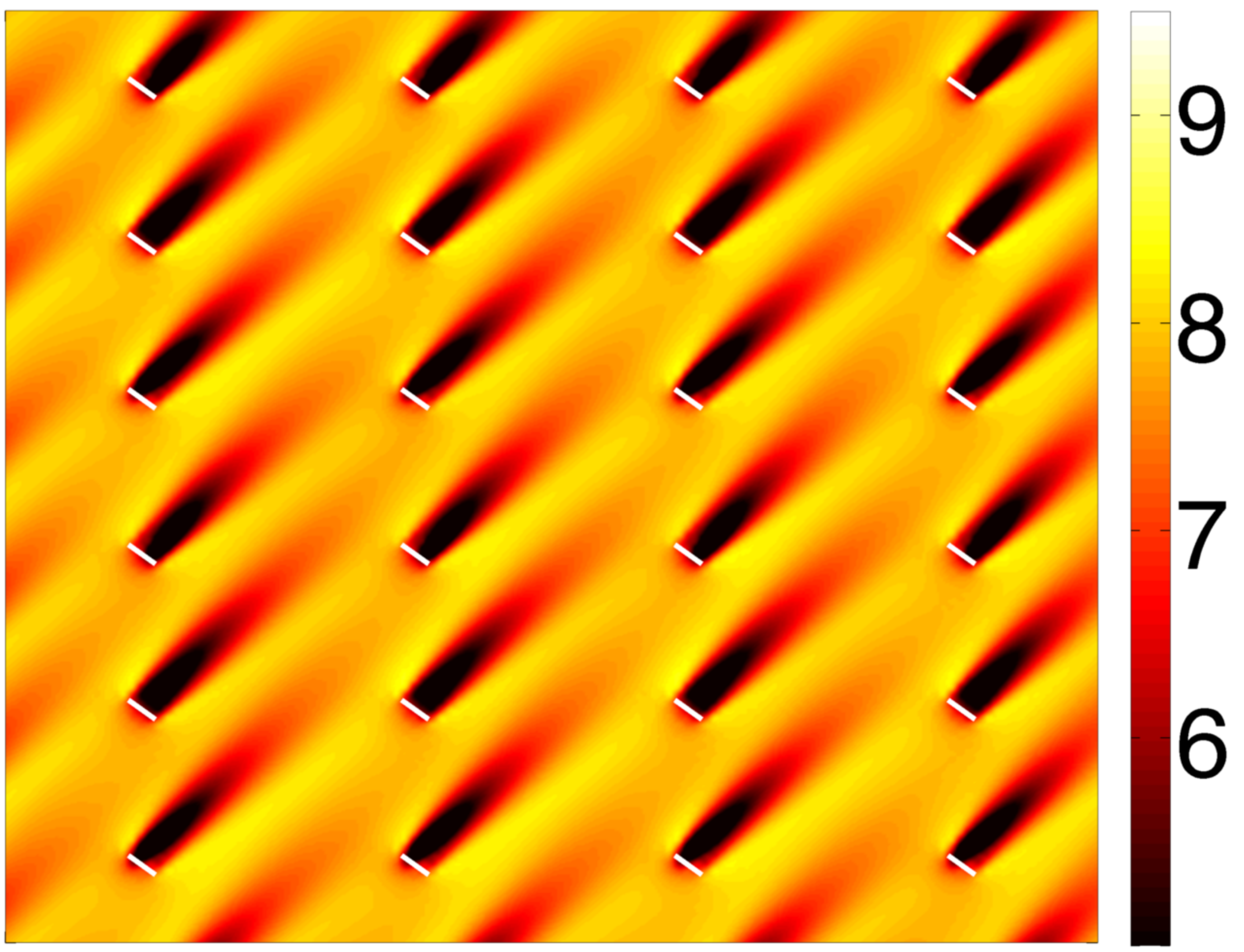}}
\subfigure[$\phi=90$ degrees]{\includegraphics[width=0.32\textwidth]{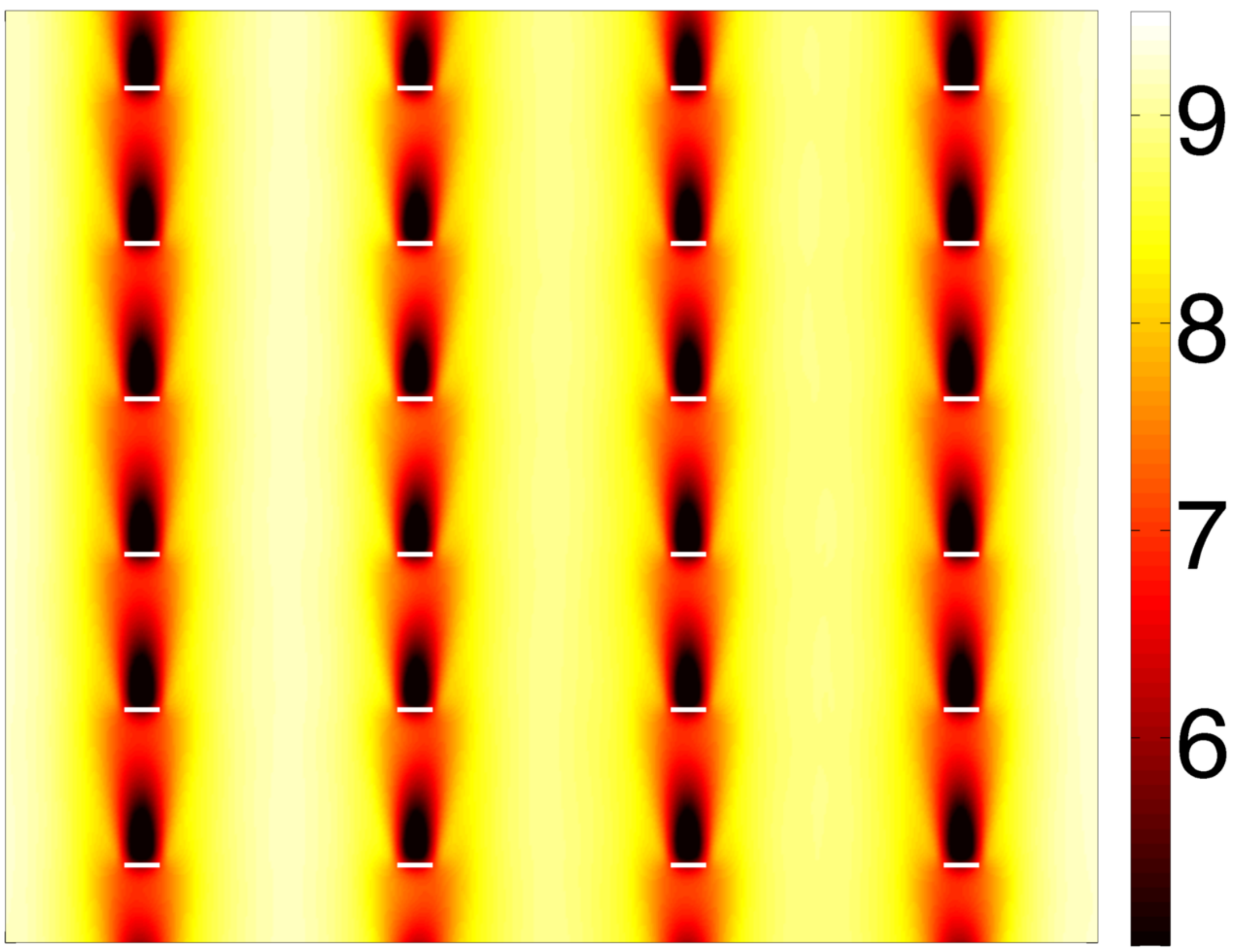}}
 \caption{Horizontally averaged velocity $\sqrt{u^2+v^2}$, where $u$ is the x-direction velocity component and $v$ is the y-direction velocity component for three mean wind-directions (a) $\phi=0$ degrees, (b) $\phi=50$ degrees, and (c) $\phi=90$ degrees. The color scale is in $m/s$.}
\label{figure11}
\end{figure}

\begin{figure}
\subfigure[]{\includegraphics[width=0.47\textwidth]{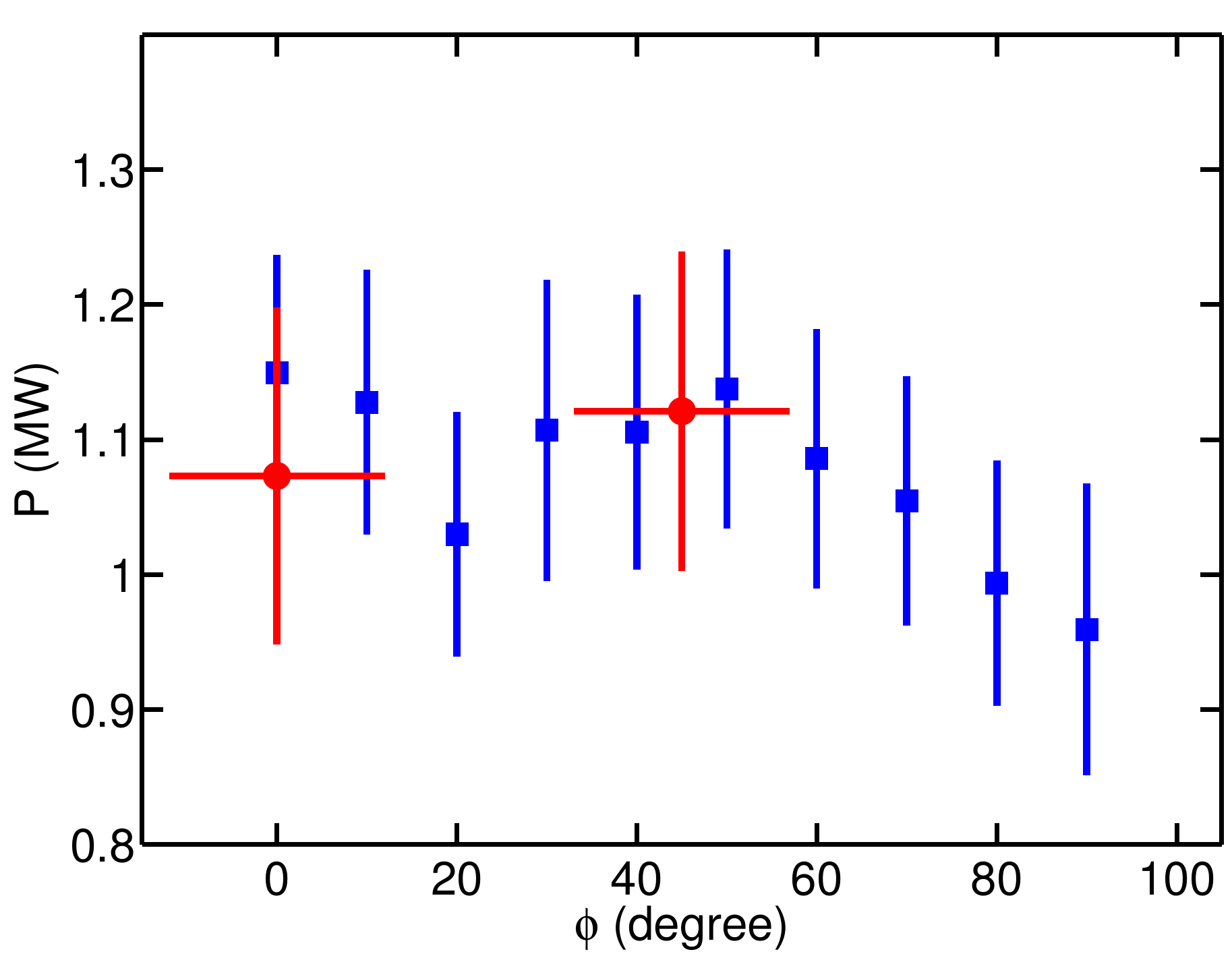}}
\subfigure[]{\includegraphics[width=0.47\textwidth]{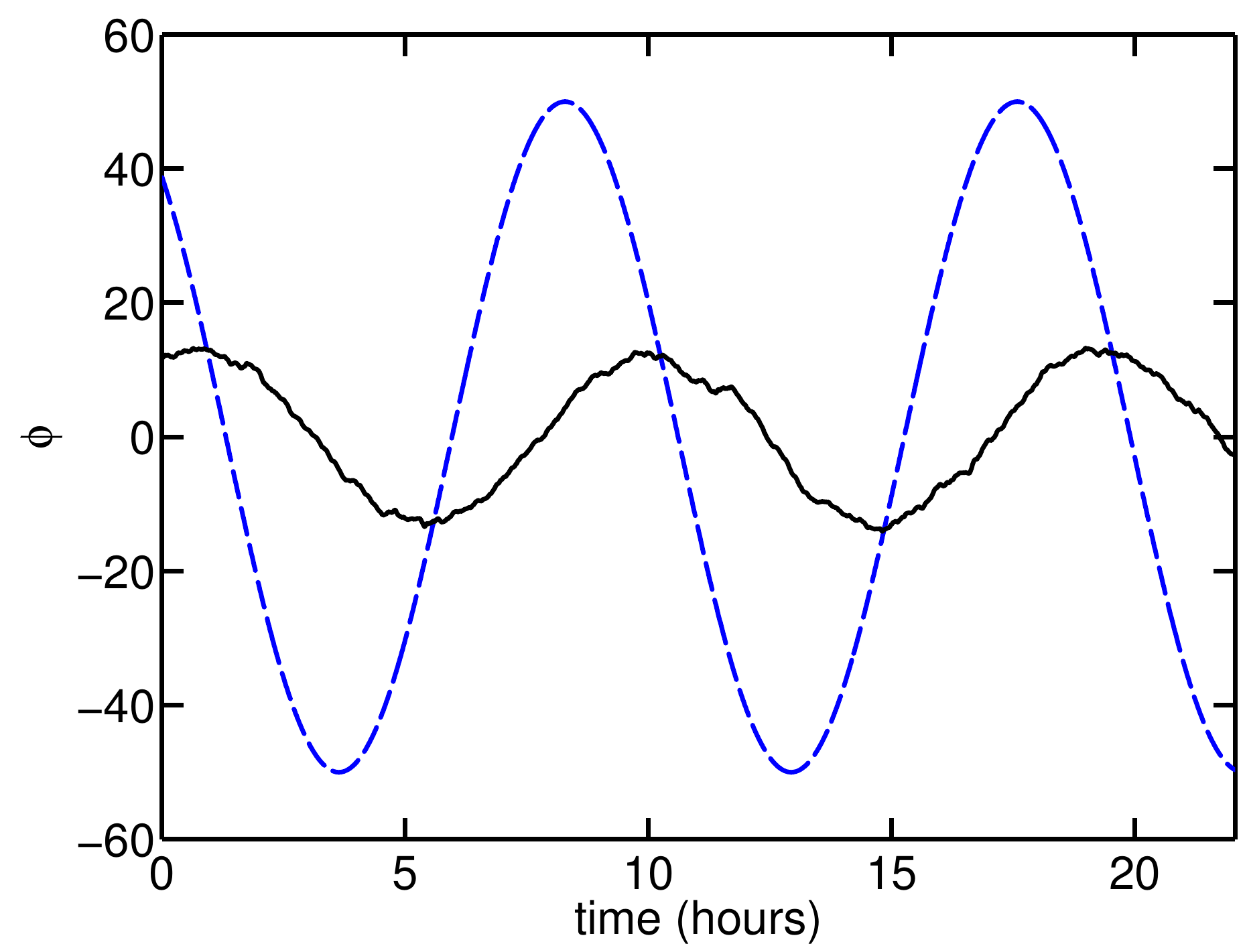}}
 \caption{(a) Average turbine power output $\langle P\rangle$ as function of imposed wind-direction. The indicated uncertainty bars correspond to the standard deviation in the power output fluctuations of all turbines. The squares indicate the results for a fixed wind-directions and the circles indicate the result for cases in which the wind-direction changes over time ($\pm 14$ degrees, indicated by the horizontal bars). (b) Comparison of the wind-direction change (solid line) measured at hub-height and the forcing direction of the mean pressure gradient (dashed line) for the simulation in which the mean wind-direction is $0$ degrees (see \S \ref{section_new3}).}
\label{figure12}
\end{figure}

To study the effect of the (stationary) wind-direction, we change $\phi$ in eq.\ \ref{eq_forcing1} and \ref{eq_forcing2}, see also table \ref{table4} and figure \ref{figure11}. Mean velocities at hub-height are shown in Fig. \ref{figure11}.  The turbine plane at which the actuator force is applied, as well as the force $f_i$ are turned to be appropriately aligned with the imposed mean flow direction. The time-averaged power is shown in figure \ref{figure12}a (squares), as function of inflow angle.  In particular we see that the wind-farm power output is lowest when $\phi\approx 90$ degrees. This corresponds to the case that the wind is aligned such that it is aligned with the short inter-turbine distances, see figure \ref{figure11}c. A somewhat higher power output is obtained when $\phi\approx0$ degrees or $\phi\approx50$ as in these cases the wind is aligned such that the inter turbine distance between turbines directly downstream is larger, see figure \ref{figure11}b. 
(Figure \ref{figure12}b and the  circles in  Fig. \ref{figure12}a are discussed below when considering time varying inflow angle.)

\begin{figure}
\subfigure[]{\includegraphics[width=0.47\textwidth]{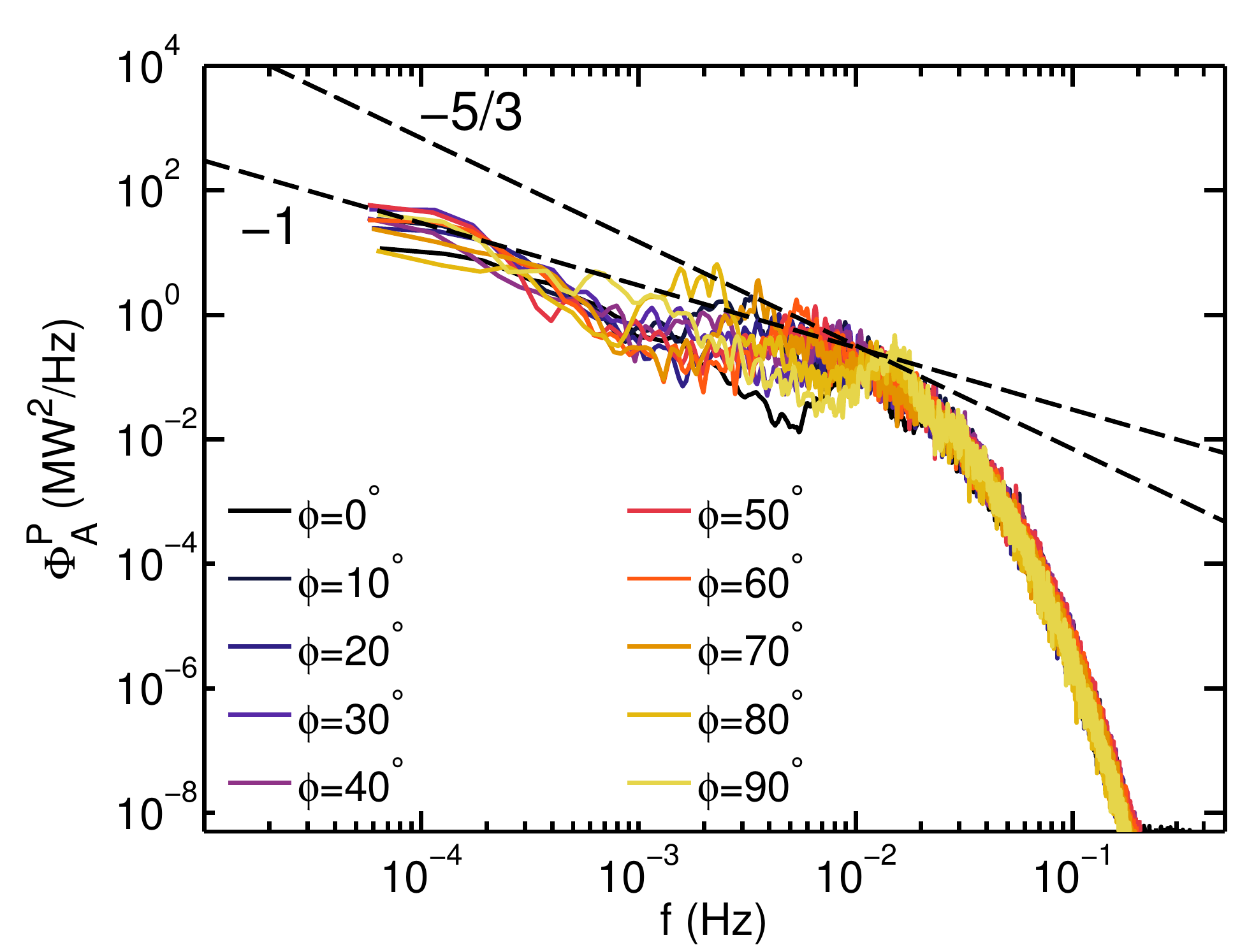}}
\subfigure[]{\includegraphics[width=0.47\textwidth]{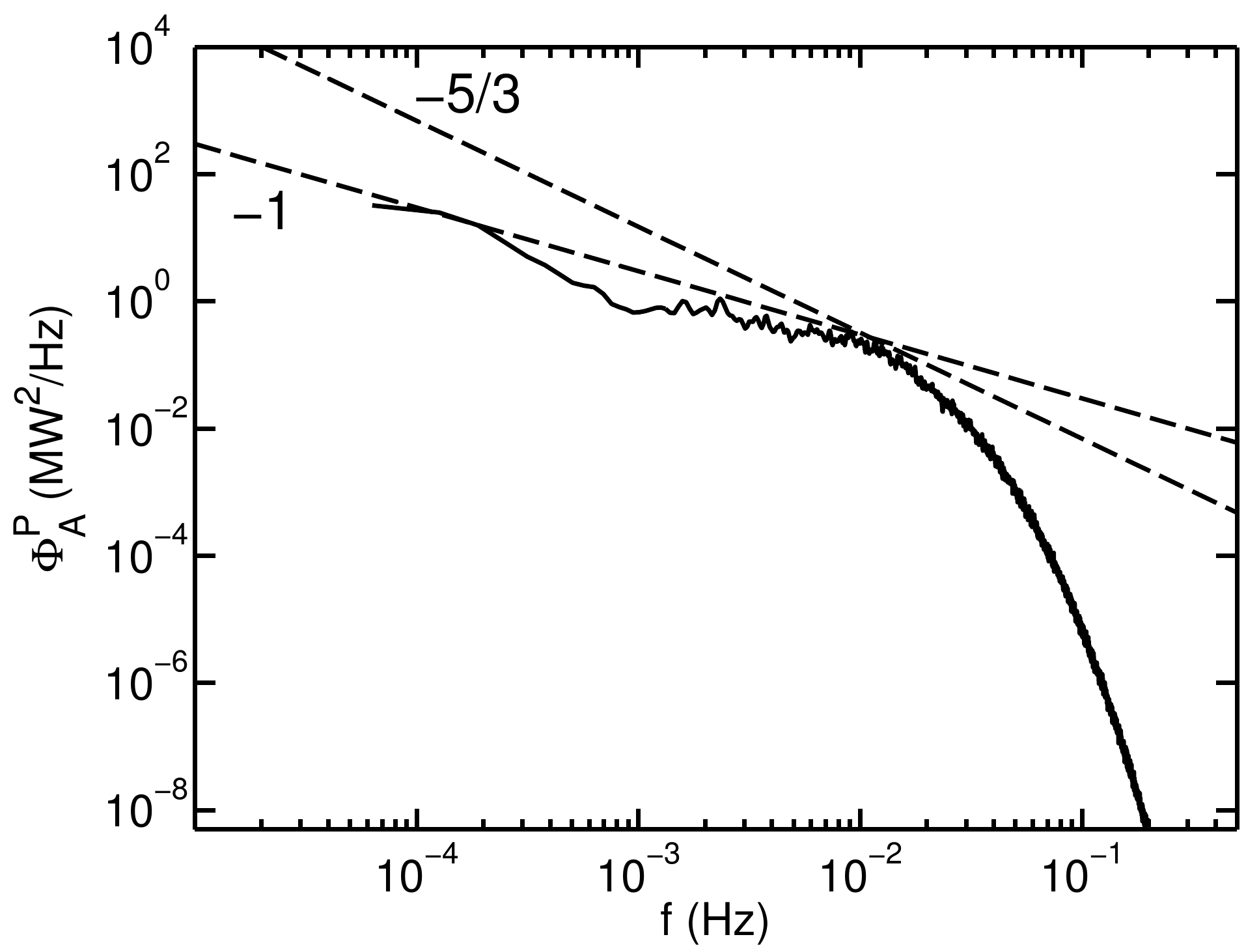}}
 \caption{(a) The spectra for the total wind-farm power output $\Phi_A^P$ for the different fixed wind-directions. Note that the position and strength of the spectral peak clearly depend on the wind-direction. (b) The average spectrum of the wind-directions shown in panel a) shows no spectral peak but a much broadened plateau.}
\label{figure13}
\end{figure}

A comparison of the power output spectra of the entire wind-farm $\Phi_A^P$ for the different cases is shown in figure \ref{figure13}a. This figure clearly reveals that the spectral peak shifts when the wind-direction is changed. Depending on whether the wind is directly aligned with the symmetry axis of the wind-farm the peak may also be more or less pronounced. If one performs an average of the spectra for all these angles (uniformly weighting each, i.e. no preferred wind direction), one obtains a complete smoothing of the spectral peak, see figure \ref{figure13}b.

\begingroup
\squeezetable
\begin{table}
\caption{Summary of the simulations of the infinite wind-farms with different wind-directions. The columns from left to right indicate the case that is considered, the stream-wise ($L_x$) and span-wise ($L_y$) size of the domain, the resolution in stream-wise ($N_x$), span-wise ($N_y$), and vertical ($N_z$) direction, the number of wind-turbines in stream-wise and span-wise direction and whether the turbines are placed in an aligned or staggered arrangement. The next five columns give the average wind-alignment $\phi$, $\mu$ indicates the amplitude of the directional change in the pressure forcing, and the average turbine power output $P$ and the standard deviation of the power output of a single wind-turbine $s_T$, and all turbines $s_A$. Statistics for rows and columns are not shown as they cannot unambiguously be defined for all these cases. The last column gives the $u_*$ value for the different cases.}
\label{table4}
\begin{center}
 \begin{tabular}{|c|c|c|c|c|c|c|c|c|c|c|c|}
 \hline
 Case	& $L_x (km)$	& $L_y (km)$	&	$N_x	 \times N_y \times N_z$	&	$N_R \times N_C$	&	$\phi$ (degrees) 	& $\mu$ (degrees)	& $\langle P\rangle [MW]$	& $s_T [MW]$ 		& $s_A [MW]$ & $u_* [m/s]$	\\ \hline
 C2		& $\pi$		& $\pi$		&	$128	\times 128	\times 128$	&	$ 8 \times 6$ 		&	0				& 0				& 1.150		& 0.499			& 0.087		& 0.913 \\ \hline	
 A2		& $\pi$		& $\pi$		&	$128	\times 128	\times 128$	&	$ 4 \times 6$ 		&	10				& 0				& 1.128		& 0.506			& 0.098		& 0.890 \\ \hline	
 A2		& $\pi$		& $\pi$		&	$128	\times 128	\times 128$	&	$ 4 \times 6$ 		&	20				& 0 				& 1.030		& 0.423			& 0.090		& 0.870 \\ \hline	
 A2		& $\pi$		& $\pi$		&	$128	\times 128	\times 128$	&	$ 4 \times 6$ 		&	30				& 0				& 1.069		& 0.484			& 0.111		& 0.893 \\ \hline	
 A2		& $\pi$		& $\pi$		&	$128	\times 128	\times 128$	&	$ 4 \times 6$ 		&	40				& 0				& 1.057		& 0.509			& 0.102		& 0.891 \\ \hline	
 A2		& $\pi$		& $\pi$		&	$128	\times 128	\times 128$	&	$ 4 \times 6$ 		&	50				& 0				& 1.377		& 0.493			& 0.103		& 0.900 \\ \hline	
 A2		& $\pi$		& $\pi$		&	$128	\times 128	\times 128$	&	$ 4 \times 6$ 		&	60				& 0				& 1.086		& 0.466			& 0.096		& 0.886 \\ \hline	
 A2		& $\pi$		& $\pi$		&	$128	\times 128	\times 128$	&	$ 4 \times 6$ 		&	70				& 0				& 1.055		& 0.436			& 0.092		& 0.874 \\ \hline	
 A2		& $\pi$		& $\pi$		&	$128	\times 128	\times 128$	&	$ 4 \times 6$ 		&	80				& 0				& 0.994		& 0.455			& 0.091		& 0.862 \\ \hline	
 A2		& $\pi$		& $\pi$		&	$128	\times 128	\times 128$	&	$ 4 \times 6$ 		&	90				& 0				& 0.959		& 0.436			& 0.108		& 0.879 \\ \hline	
 A2		& $\pi$		& $\pi$		&	$128	\times 128	\times 128$	&	$ 4 \times 6$ 		&	0				& 50				& 1.073		& 0.482			& 0.125		& 0.968 \\ \hline	
 A2		& $\pi$		& $\pi$		&	$128	\times 128	\times 128$	&	$ 4 \times 6$ 		&	45				& 50				& 1.121		& 0.485			& 0.118		& 0.987 \\ \hline	
 \end{tabular}
 \end{center}
\end{table}
\endgroup

\subsection{Time-varying wind-direction} \label{section_new3}
\begin{figure}
\subfigure[]{\includegraphics[width=0.47\textwidth]{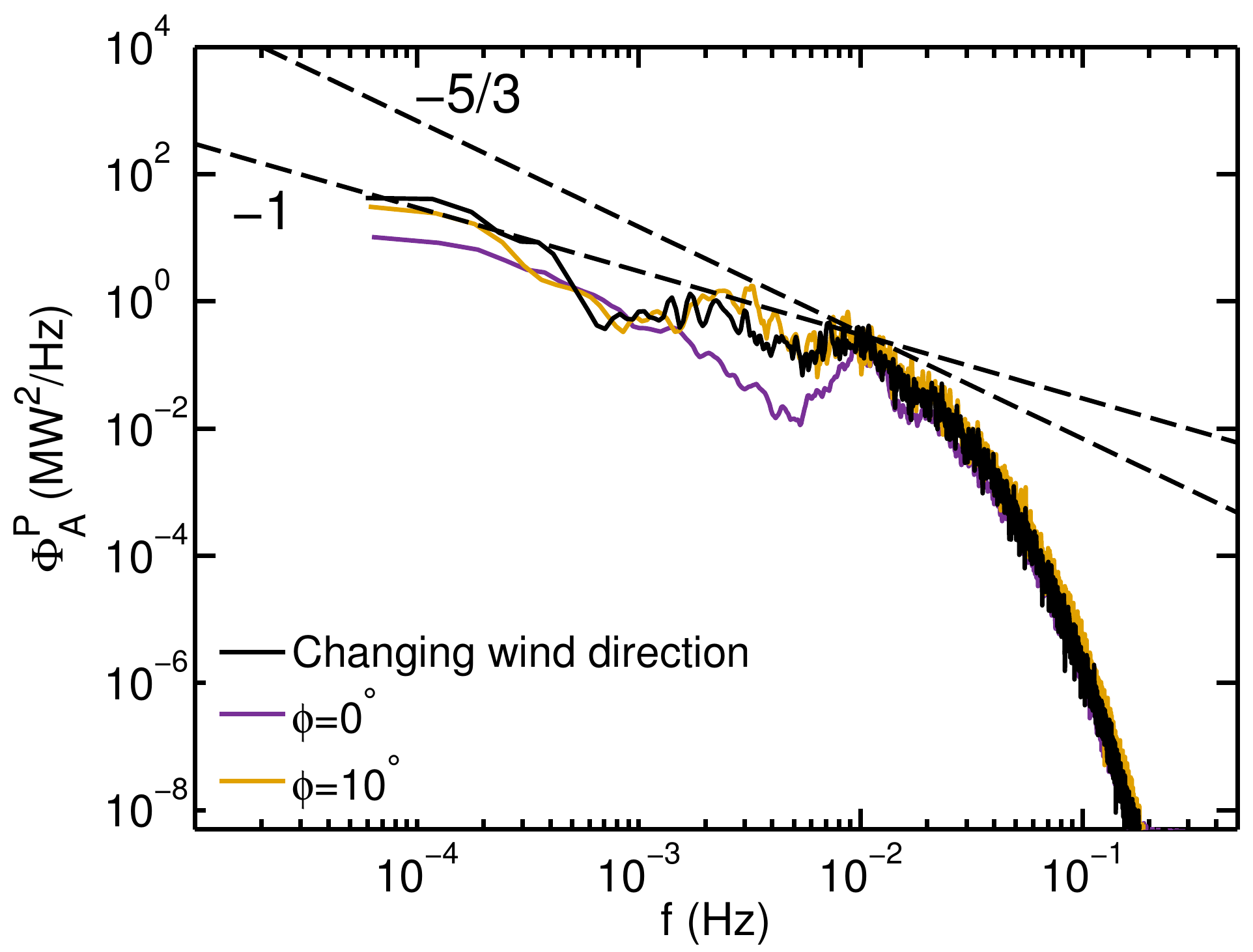}}
\subfigure[]{\includegraphics[width=0.47\textwidth]{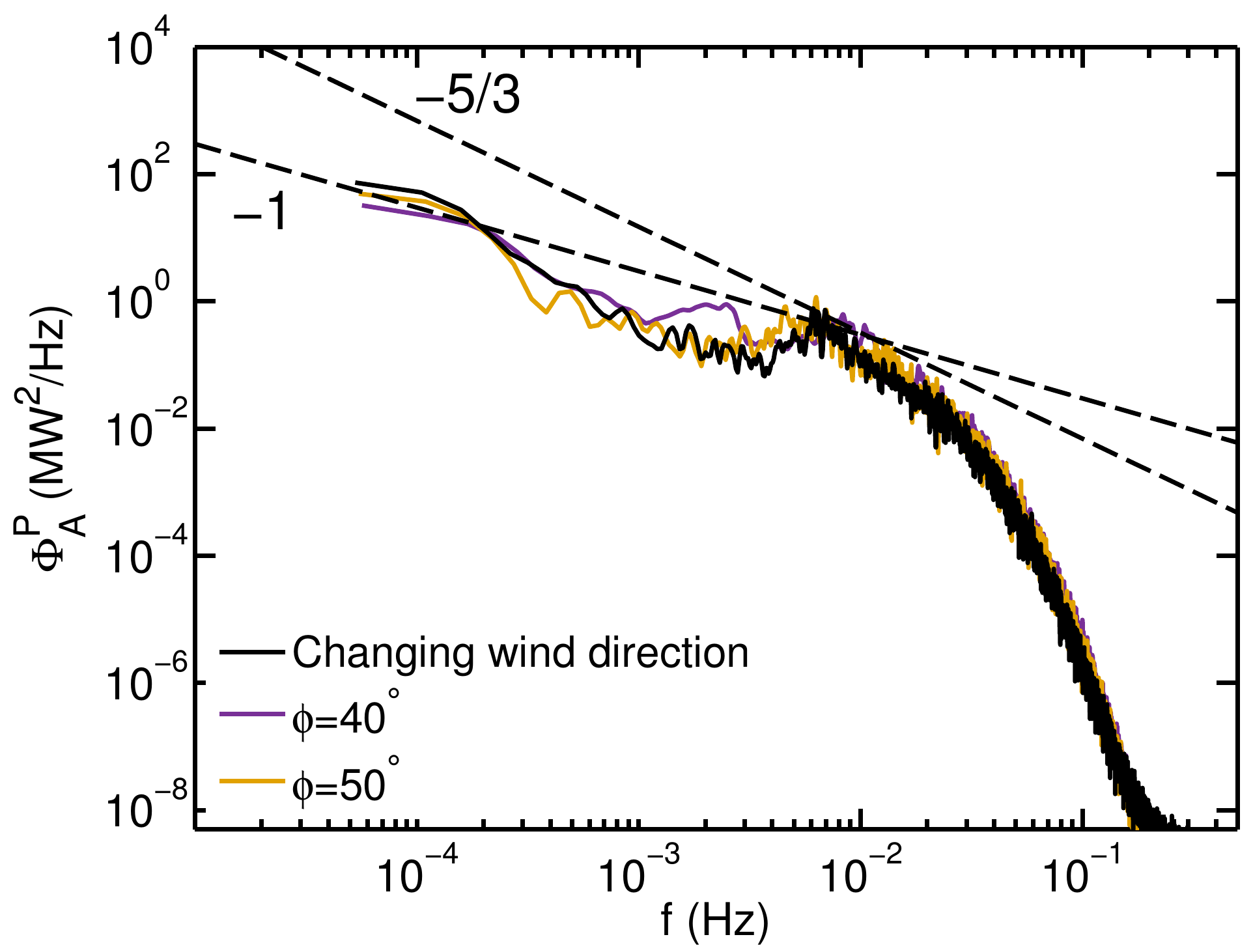}}
 \caption{(a) Spectra for changing wind-direction ($\pm 14$ degrees) around the $0$ degree alignment (b) spectra for changing wind-direction ($\pm 14$ degrees) around the $45$ degree wind-alignment. In each panel the spectra for wind-directions in the regime in which the wind-direction is changed are given as reference.}
\label{figure14}
\end{figure}

To model the effect of a changing wind-direction we set $\mu=50$ degrees in eq.\ \ref{eq_forcing1} and \ref{eq_forcing2}, and we do this for two main wind-directions, i.e. $\phi=0$ and $\phi=45$ degrees. Figure \ref{figure12}b shows that this change in the direction of the pressure forcing leads to changes in the mean wind-direction of $\pm$ 14 degrees at hub height. In order to make sure that the turbines remain aligned with the mean wind-direction, the actuator force disk is turned during the simulation. In figure \ref{figure12}a we can see that the average total wind-farm power output obtained from the simulations with the dynamically changing wind-direction are in good agreement with the results obtained for the different fixed wind-direction cases considered before. In figure \ref{figure14} we show the spectra for the total wind-farm power output for these two cases. Figure \ref{figure14} shows that the spectral peak is smoothed when the wind-direction changes over time. How much the peak is smoothed depends on the mean wind-direction. For the $0$ degree mean wind-direction the smoothing due to the changing wind-direction is relatively large. We believe this is observed because due to the constantly changing wind-direction the wind is almost never perfectly aligned with the turbines, which leads to the strong peak for the $0$ degree wind case and is significantly less with a small $\phi$. For the case in which the wind-direction fluctuates around the  $45$ degree orientation the wind is never really aligned with the turbines and therefore the peak position does not change much, which is indicated by the reference spectra in figure \ref{figure14}b.

\subsection{Finite size wind-farms} \label{section4b}
To ascertain if some of the effects observed before (spectral peak, -5/3 power-law, etc.), are due to the fact that we consider fully periodic boundary condition in an
infinite wind farm, here we consider wind farms with an entrance region. LES of finite size wind-park are performed using the concurrent-precursor inflow LES method described in Ref.\ \cite{ste14,ste14b,ste14c}. A precursor LES is performed in one domain and is used as inflow for the simulation domain with the wind farm. Hence the first row of wind turbines is exposed to an unperturbed fully developed (neutral) atmospheric boundary layer. 

In order to better capture the development of the initial internal boundary layer without any effects from the top boundary, we choose to perform simulations of   finite size wind-farm using a domain height of $L_z=2$km instead of $1$ km.  Also the stream-wise number of grid points is significantly higher due to the use of the concurrent precursor method we use to generate the inflow conditions \cite{ste14}. Hence this simulation is more costly and thus only a single case is considered for this comparison study. 
Table \ref{table3} shows the details of the turbine power outputs from row 4 to row 12 of our finite size wind-farm.  As for the aligned configuration the power output is approximately independent of the stream-wise distance in the wind-farm \cite{ste14}. This fact allows us to compare the results with the infinite wind-farm simulations presented before. The number of turbines in this part of the wind-farm is equal to the number of turbines used in simulation C2, see table \ref{table1}, and we find a very good agreement between both cases for the average power output and the standard deviations of power output of the different aggregates. 

\begin{figure}
\subfigure[]{\includegraphics[width=0.47\textwidth]{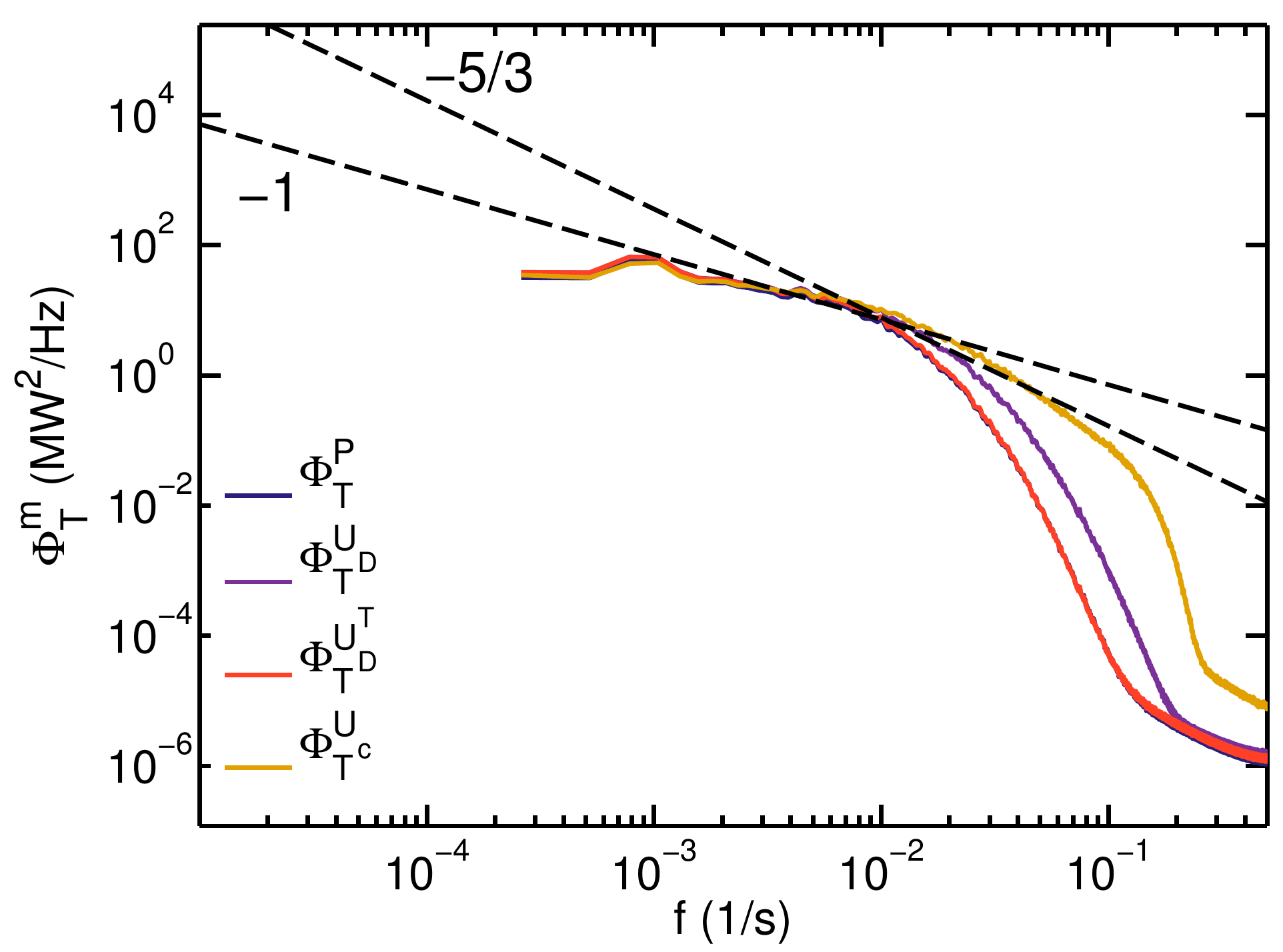}}
\subfigure[]{\includegraphics[width=0.47\textwidth]{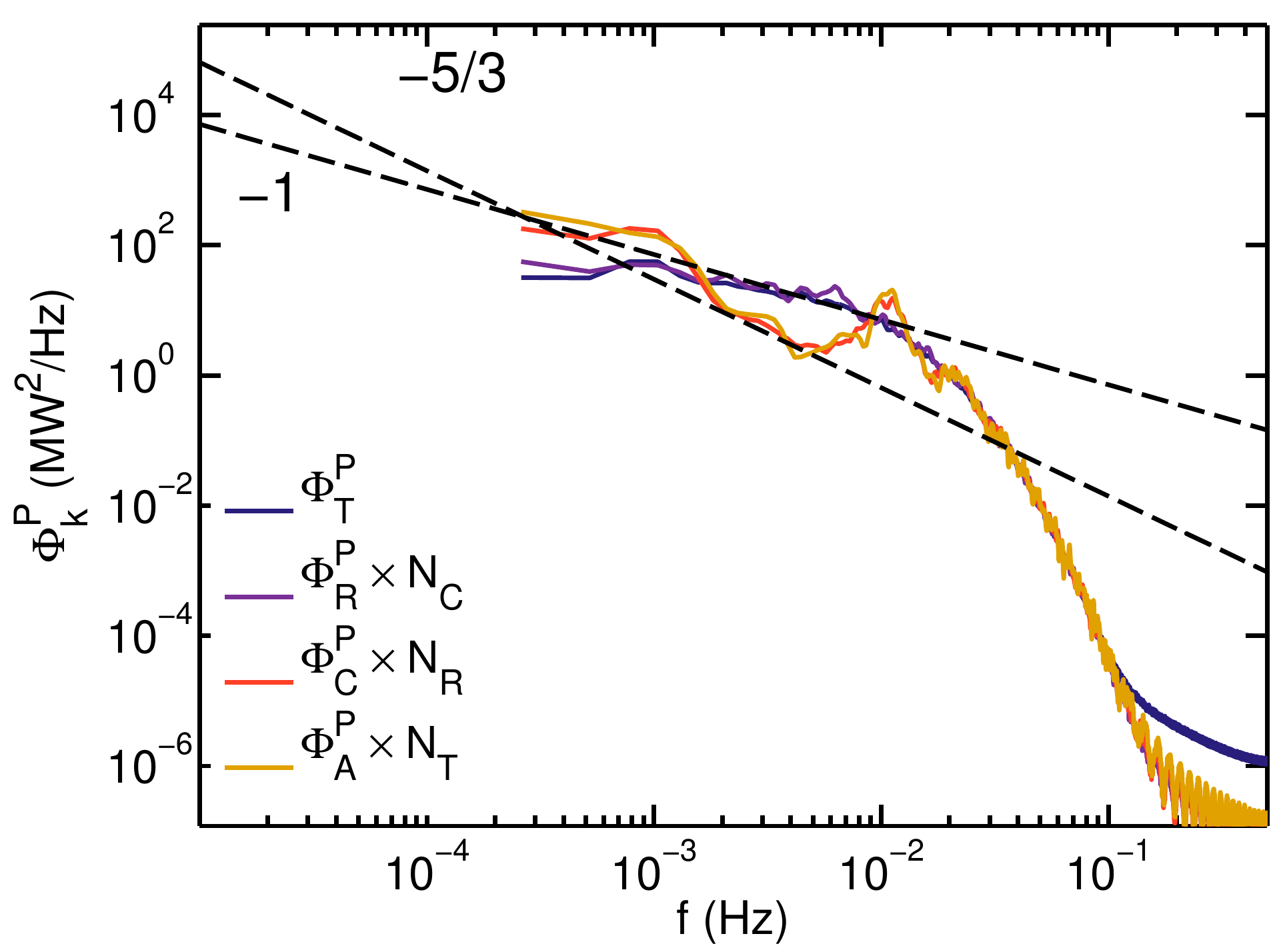}}
 \caption{a) Spectra for the fluctuations of velocity averaged over the turbine surface $\Phi_T^{U_D}$, the time averaged velocity averaged over the turbine surface $\Phi_T^{U_D^T}$, the power of the turbine $\Phi_T^P$, and the spectrum for the stream-wise velocity at the turbine center $\Phi_T^{U_C}$ for a finite size wind-farm (simulation C2f) look similar to the ones obtained for the infinite wind-farms, see figure \ref{figure4}b. Also here $\Phi_T^{U_C}$, $\Phi_T^{U_D}$, and $\Phi_T^{U_D^T}$ are arbitrarily shifted to allow comparison of the shape of the $\Phi_T^{P}$ spectrum. b) Comparison of the spectra of the power for a single turbine $\Phi_T^{P}$, a row of turbines $\Phi_T^{R}$ (in the span-wise flow direction), a column of turbines $\Phi_T^{C}$ (in the stream-wise flow direction), and the entire wind-farm $\Phi_T^{A}$ for a finite size wind-farm (simulation C2f). Note that the observed features are very similar as for an infinite wind-farm, see figure \ref{figure6}b.}
\label{figure15}
\end{figure}

Figure \ref{figure15}a shows that the spectra of velocity and power obtained for a single turbine is similar as we observed for a single turbine in an infinite wind-park, compare with figure \ref{figure4}b. In figure \ref{figure15}b the power output for the finite wind-farm is considered in the same way as was done for the infinite wind-farm in figure \ref{figure8}b, again using the data from rows 4 to 12 only. It is important to note that the main characteristic observed in the simulations of this finite size wind-farm are similar to the results we obtained for the infinite wind-farms. This verifies that the larger than expected reduction of the aggregate spectrum at the intermediate frequencies is an effect of the complex interaction between the turbines placed in the stream-wise direction and not caused by the periodic boundary conditions used in the simulations of the infinite wind-farms.

\begingroup
\squeezetable
\begin{table} 
\caption{Same as table \ref{table1}, but now for the finite size wind-farm simulation considered based on the data of the fourth to twelfth turbine row. Note that a good agreement with case $C2$, the closest case to this finite size wind-farm case, is obtained.}
\label{table3}
\begin{center}
 \begin{tabular}{|c|c|c|c|c|c|c|c|c|c|c|c|}
 \hline
 Case	& $L_x (km)$	& $L_y (km)$	&	$N_x	 \times N_y \times N_z$	&	$N_t$		&	Positioning 	& $\langle P\rangle [MW]$	& $s_T [MW]$ 	& 	$s_A [MW]$	&$s_R [MW]$	& $s_C [MW]$	 &	$u_* [m/s]$	\\ \hline
 C2f		& $4\pi$		& $\pi$		&	$512\times 128	\times 256$	&	$ 13 \times 6$	&	aligned		& 1.186		&	0.490	& 	0.086		& 0.193 		& 0.240		 &	0.542\\
 \hline
 \end{tabular}
\end{center}
\end{table}
\endgroup

\section{Discussion and conclusions} \label{section5}
In this paper we have analyzed the characteristics of the power fluctuations that are observed in large eddy simulations (LES) of extended wind-parks in an atmospheric boundary layer. The fluctuations correspond to the power extracted from the flow from microscale atmospheric turbulence fluctuations. We consider various aggregates of wind-turbines such as the total average power signal, or sub-averages within the wind-farm. We find that the power variations of the total wind-park in the intermediate to low frequency range decrease more than one would expect if one assumes the power fluctuations at the turbines to be uncorrelated among themselves. This effect is caused by complex interactions between turbines placed downstream from each other, yielding non-trivial collective phenomena.  We also observe that the frequency spectra of the total wind-farm output show a decay that follows approximately a $-5/3$ power-law scaling regime, qualitatively consistent with similar observations made in field-scale operational wind-parks by Apt \cite{apt07}. 

At first one may have expected that the $-5/3$ scaling is somehow due to a Kolmogorov scaling of the underlying turbulent wind-speed. Using the arguments that power is proportional to the cube of velocity, and using a decomposition of the velocity into its mean and fluctuations as $u^3 = (\overline{u}+u^\prime)^3$ and assuming that the fluctuations are small compared to the mean velocity (typically one has turbulence intensities in the stream-wise velocity of about $10 \%$), one can write $u^3 \sim \overline{u}^3$ + 3 $\overline{u}^2 u^\prime$+..  Therefore this argument suggests that to first order the fluctuations in power output should be proportional to those of $u^\prime$, i.e.\ possibly yielding a $-5/3$ spectrum. However, clearly the present results demonstrate that such a scaling is obtained only after averaging particular turbines, in frequency ranges in which the velocity itself did not display a $-5/3$ spectrum. Therefore, the emergence of a $-5/3$ scaling is possibly due to other effects, for which  we do not have plausible explanations so far. 

We find that as long as the wind-direction is fixed changes in the wind-speed do not really influence this effect assuming that they remain operating in regime II. However, our simulations show that the effect is strongest when the wind-direction is directly aligned with the wind-farm columns. For other wind-directions the spectral peak is less pronounced. Also changes in the wind-direction over time tend to smooth the spectral peak, so the effect is not necessarily as clearly observable in data from operational wind-farms as it can be observed in LES with an imposed constant mean velocity direction. Better understanding of the origins of the -5/3 power law thus remains an open challenge. In addition, it would be very interesting to explore the spectral characteristics of power fluctuations as function of atmospheric stability and design parameters of the wind-park, such as the turbine spacing and the hub height of the turbines. Possible effects stemming from the coupling of wind-turbine arrays  to power network are also worth studying in future work.

{\it Acknowledgements}: The authors thank Dr.\ Dennice Gayme, Dr.\ Jason Graham and Claire VerHulst for interesting conversations and comments. They also acknowledge Claire VerHulst's contribution with the coding of the actuator-disk model in the LES code, and Jason Graham's work on developing and testing the concurrent precursor simulation method. RJAMS's work is supported by the program `Fellowships for Young Energy Scientists' (YES!) of the Foundation for Fundamental Research on Matter (FOM), which is financially supported by the Netherlands Organization for Scientific Research (NWO). Further partial support has been provided by the US National Science Foundation from grants NSF-CBET 1133800 and OISE 1243482. The computations have been performed on a local cluster at Johns Hopkins and on the LISA cluster of SARA in the Netherlands.

\end{document}